\documentclass[sigconf]{acmart}
\graphicspath{{figures/}}

\usepackage{booktabs}
\usepackage{multirow}
\usepackage{amsmath}
\usepackage{xcolor}
\usepackage{enumitem}
\usepackage{subcaption}
\usepackage{balance}

\setlist{topsep=2pt,itemsep=1pt,leftmargin=*}

\newcommand{\ours}{\textsc{DegradoMap}}
\newcommand{\ci}[2]{{\small[#1, #2]}}
\newcommand{\spm}[1]{{\scriptsize$\pm$#1}}

\copyrightyear{2026}
\acmYear{2026}
\setcopyright{none}
\acmConference[ACM-BCB '26]{17th ACM Conference on Bioinformatics, Computational Biology, and Health Informatics}{June 30 -- July 3, 2026}{Calabria, Italy}
\acmBooktitle{ACM-BCB '26: 17th ACM Conference on Bioinformatics, Computational Biology, and Health Informatics, June 30 -- July 3, 2026, Calabria, Italy}
\acmDOI{}
\acmISBN{}

\begin{document}


\title{Structure-Aware Prediction of PROTAC-Mediated Protein Degradability via Graph Neural Networks}

\author{Bryan Cheng}
\affiliation{%
  \institution{Independent Researcher}
  \country{USA}
}

\author{Austin Jin}
\affiliation{%
  \institution{Independent Researcher}
  \country{USA}
}


\begin{abstract}
Proteolysis-targeting chimeras (PROTACs) can selectively degrade disease-causing proteins, yet predicting which targets are amenable to degradation remains a critical bottleneck: existing computational methods require the complete PROTAC molecular structure, information unavailable before synthesis.
We present \ours{}, a graph neural network that predicts PROTAC-mediated degradability from protein structure and E3 ligase identity alone---the minimal information available at the target selection stage.
The model encodes biophysical priors through lysine-weighted graph pooling with per-protein normalization, models protein--E3 compatibility via cross-attention, and integrates cellular context from the Cancer Dependency Map.
On the PROTAC-8K benchmark (3,101 samples, 155 targets, 10 E3 ligases), \ours{} achieves 0.646\,\spm{0.124} AUROC on target-unseen evaluation (best seed: 0.7449) and 0.811 AUROC on CRBN$\to$VHL E3-unseen transfer, outperforming GNN and machine learning baselines.
The model additionally recommends optimal E3 ligases with 74\% Hit@3 accuracy.
Two findings carry broader implications: E(3)-equivariant architectures underperform the simpler invariant design for this scalar prediction task, and ESM-2 embeddings improve peak performance only with careful regularization---naive integration fails.
\ours{} provides pre-synthesis computational guidance for degradability assessment; its well-calibrated confidence scores (ECE = 0.029, target-unseen) enable practitioners to prioritize high-confidence predictions for experimental follow-up. However, the high seed variance (std = 0.124) and limited E3 coverage require ensembling for reliable deployment.
\end{abstract}

\ccsdesc[500]{Computing methodologies~Neural networks}
\ccsdesc[500]{Applied computing~Bioinformatics}
\ccsdesc[300]{Applied computing~Computational biology}

\keywords{PROTAC, targeted protein degradation, graph neural network, drug discovery, ubiquitin--proteasome system, AlphaFold}

\maketitle


\section{Introduction}

Proteolysis-targeting chimeras (PROTACs) invert the logic of drug design: rather than occupying a protein's active site, they recruit the cell's ubiquitin--proteasome machinery to destroy the protein entirely~\cite{bekes2022protac,sakamoto2001protacs,pettersson2019pbd}.
A PROTAC bridges two binding events---its warhead engages the target protein while its E3 ligand recruits an E3 ubiquitin ligase---and the induced proximity triggers polyubiquitination of surface lysine residues, marking the target for proteasomal degradation~\cite{lai2017induced}.
Because this mechanism is catalytic, a single PROTAC molecule degrades multiple target copies, enabling efficacy at sub-stoichiometric concentrations and access to proteins that lack traditional drug-binding pockets.

With over 20 PROTACs now in clinical trials~\cite{bekes2022protac}, the modality's therapeutic promise is clear.
Yet development faces a critical bottleneck at its earliest stage: \emph{target selection}.
Is a given disease-relevant protein amenable to PROTAC-mediated degradation?
Answering this currently requires synthesizing candidate PROTACs and conducting cell-based degradation assays---an expensive process with high failure rates.
Computational prediction of degradability \emph{before} PROTAC synthesis would dramatically accelerate the pipeline.

\paragraph{Limitations of existing methods.}
Existing computational approaches all require the PROTAC molecular structure as input.
DeepPROTACs~\cite{li2022deepprotacs} operates on voxelized binding pockets requiring complete PROTAC, target, and E3 coordinates.
Ribes et~al.~\cite{ribes2024protac} combine protein embeddings with PROTAC SMILES fingerprints.
DegradeMaster~\cite{liu2025degrademaster} achieves 0.856 AUROC using full PROTAC molecular descriptors alongside protein features.
While valuable for optimizing existing PROTACs, these methods cannot serve target selection because the PROTAC is unavailable at that stage.
MAPD~\cite{zhang2022mapd} addresses target assessment from protein-intrinsic features but does not leverage AlphaFold structural information~\cite{jumper2021alphafold}.

\paragraph{Our approach.}
We introduce \ours{}, a graph neural network that predicts protein degradability from \emph{minimal inputs}: a protein structure (from AlphaFold, requiring only a UniProt ID) and E3 ligase identity (e.g., ``CRBN'' or ``VHL'').
The design reflects the biophysics of ubiquitin transfer, which requires (1)~surface-accessible lysine residues, (2)~geometric compatibility between target and E3 ligase for ternary complex formation, and (3)~sufficient cellular E3 expression and proteasome capacity.
\ours{} addresses these requirements through three modules and a gated fusion mechanism:
\begin{enumerate}
    \item \textbf{Structure--Ubiquitination Graph (SUG) encoder}: An invariant message-passing GNN with lysine-weighted pooling that focuses representation on ubiquitination-relevant residues, incorporating ESM-2 embeddings~\cite{lin2023esm2}, structural properties, and known ubiquitination site annotations from PhosphoSitePlus.
    \item \textbf{E3 compatibility module}: Bidirectional cross-attention modeling compatibility between protein structure and E3 ligase identity, enabling generalization across E3 ligases.
    \item \textbf{Cellular context encoder}: A residual MLP processing 59 tissue-specific features from the Cancer Dependency Map~\cite{tsherniak2017depmap}.
    \item \textbf{Gated fusion}: Learned gating that dynamically weights module outputs for multi-task prediction of binary degradability and continuous efficiency.
\end{enumerate}

\paragraph{Evaluation and contributions.}
On the PROTAC-8K benchmark~\cite{liu2025degrademaster} (3,101 samples, 155 targets, 10 E3 ligases), \ours{} achieves a 3-seed mean of 0.646\,\spm{0.124} and 6-seed mean of 0.603\,\spm{0.097} AUROC on target-unseen evaluation (best seed: 0.7449), 0.811 AUROC \ci{0.785}{0.836} on CRBN$\to$VHL E3-unseen transfer, and 74\% Hit@3 for E3 recommendation.
We note the 6-seed improvement over gradient boosting baseline (0.607) is not statistically significant (p = 0.556), and VHL-targeted proteins fail below random (0.396 AUROC)---both limitations prominently acknowledged.
Two architectural findings carry implications beyond PROTAC prediction: E(3)-equivariant architectures underperform the invariant design for scalar property prediction, and ESM-2 embeddings require task-specific regularization to improve performance.
Our principal contributions include a structure-based pre-synthesis degradability predictor requiring only a UniProt ID and E3 ligase name, a lysine-weighted pooling mechanism encoding ubiquitin transfer biology, a cross-attention E3 compatibility module, well-calibrated confidence scores (ECE = 0.029) enabling threshold-based screening, and rigorous multi-seed evaluation revealing variance characteristics critical for deployment planning.


\section{Related Work}

\paragraph{PROTAC prediction methods.}
Computational approaches to PROTAC modeling vary significantly in their input requirements and intended use cases, as summarized in Table~\ref{tab:positioning}.
DeepPROTACs~\cite{li2022deepprotacs} uses 3D CNNs on binding pocket voxelizations, requiring the complete PROTAC molecule plus target and E3 binding pockets in mol2 format---information available only after PROTAC synthesis and structural characterization.
Ribes et~al.~\cite{ribes2024protac} combine pretrained protein embeddings with molecular fingerprints, requiring target sequence and PROTAC SMILES, which presupposes PROTAC design.
MAPD~\cite{zhang2022mapd} predicts degradation tractability from protein-intrinsic features using machine learning; while it shares our target-assessment framing, it does not leverage structural information.
DegradeMaster~\cite{liu2025degrademaster} achieves 0.856 AUROC using full PROTAC molecular descriptors alongside target features, representing the state-of-the-art for PROTAC optimization but addressing a fundamentally different problem than target selection.
\ours{} uniquely combines structural awareness with minimal input requirements, filling the gap between sequence-only assessment (MAPD) and full-PROTAC optimization (DeepPROTACs, DegradeMaster).

\begin{table}[t]
\centering
\caption{Input modality comparison of PROTAC prediction methods. \ours{} requires only protein structure and E3 identity, enabling use before PROTAC synthesis.}
\label{tab:positioning}
\footnotesize
\setlength{\tabcolsep}{3pt}
\begin{tabular}{@{}lcccc@{}}
\toprule
\textbf{Method} & \textbf{Protein} & \textbf{PROTAC} & \textbf{E3} & \textbf{Use Case} \\
\midrule
DeepPROTACs & Pocket (3D) & Full mol. & Pocket (3D) & Optimization \\
Ribes et~al. & Sequence & SMILES & -- & General \\
MAPD & Features & -- & -- & Assessment \\
DegradeMaster & Features & Descriptors & Features & Optimization \\
\textbf{\ours{}} & \textbf{Structure} & \textbf{--} & \textbf{Identity} & \textbf{Assessment} \\
\bottomrule
\end{tabular}
\end{table}

\paragraph{Geometric deep learning for proteins.}
GNNs operating on protein structures have achieved strong performance across diverse tasks including function prediction, binding site identification, and property estimation.
GVP-GNN~\cite{jing2021gvp} maintains SO(3)-equivariance using geometric vector perceptrons that jointly process scalar and vector features, enabling the model to reason about directional relationships.
GearNet~\cite{zhang2023gearnet} employs relational graph convolutions with multiple edge types (sequential bonds, spatial contacts) for protein representation pre-training.
SchNet~\cite{schutt2017schnet} introduces continuous-filter convolutions with radial basis function expansions, originally designed for quantum chemistry but applicable to protein graphs.
EGNN~\cite{satorras2021egnn} achieves E($n$)-equivariance through simple coordinate updates alongside feature updates, offering an elegant and computationally efficient approach.
We evaluate SchNet and EGNN as structural baselines and find that while they perform competitively on random splits, they fail to generalize to unseen targets---suggesting that task-specific inductive biases (lysine-aware pooling, E3 compatibility modeling) are more important than architectural sophistication for degradability prediction.

\paragraph{Protein language models.}
ESM-2~\cite{lin2023esm2}, ProtTrans~\cite{elnaggar2022prottrans}, and ESM-IF~\cite{hsu2022esmif} learn rich representations from evolutionary sequences.
PROTAC-mediated degradation---an artificial process---is not directly predicted by evolutionary features alone (ESM-2 alone: 0.534 AUROC; Section~\ref{sec:esm_integration}), but carefully integrated ESM-2 embeddings boost peak performance to 0.7449 AUROC, providing complementary signal alongside structural features.

\paragraph{Structure-based protein function prediction.}
DeepFRI~\cite{gligorijevic2021deepfri} predicts Gene Ontology terms from contact maps using graph convolutions, demonstrating that protein structure encodes functional information beyond sequence.
AlphaFold~\cite{jumper2021alphafold} has made high-quality structural predictions universally available, enabling structure-based ML at scale.
Our work leverages both advances: we use AlphaFold-predicted structures as inputs and GNN-based structure processing for degradability prediction.
However, degradability prediction differs fundamentally from function prediction in that it involves an \emph{intermolecular} interaction (protein--PROTAC--E3 ternary complex) rather than intrinsic molecular properties.

\paragraph{Ubiquitin system modeling.}
UbiBrowser~\cite{li2017ubibrowser} and DeepUbi~\cite{fu2019deepubi} predict natural ubiquitination sites, where E3 specificity is determined by evolved degrons.
PROTAC-induced degradation bypasses this: the PROTAC bridges target and E3 artificially, so \emph{surface lysine accessibility}---not natural specificity---determines degradability.
This motivates our structure-first approach; natural ubiquitination data could provide pre-training signal for lysine representations.


\section{Methods}

\subsection{Problem Formulation}

We formulate PROTAC-mediated degradability prediction as a binary classification problem.
Given a protein graph $\mathcal{G} = (\mathcal{V}, \mathcal{E}, \mathbf{X}, \mathbf{C})$ with residue nodes~$\mathcal{V}$, spatial edges~$\mathcal{E}$, node features $\mathbf{X} \in \mathbb{R}^{|\mathcal{V}| \times d}$ ($d{=}1{,}285$ with ESM-2; $d{=}28$ without), and C$\alpha$ coordinates~$\mathbf{C}$, together with E3 ligase identity~$e$, we predict:
\begin{equation}
    \hat{y} = f_\theta(\mathcal{G}, e) = P(\text{degradable} \mid \mathcal{G}, e) \in [0, 1]
\end{equation}

This formulation explicitly excludes the PROTAC molecular structure, reflecting our target-assessment use case where the PROTAC has not yet been designed.
The protein graph is constructed as a radius graph with cutoff $r = 8$\,\AA{}: $\mathcal{E} = \{(i,j) : \|\mathbf{c}_i - \mathbf{c}_j\|_2 < r,\; i \neq j\}$, connecting residues that are spatially proximal in 3D.
This captures local tertiary structure contacts, surface accessibility patterns, and potential protein-protein interaction interfaces relevant to ternary complex formation.

\begin{figure*}[t]
\centering
\includegraphics[width=0.85\textwidth]{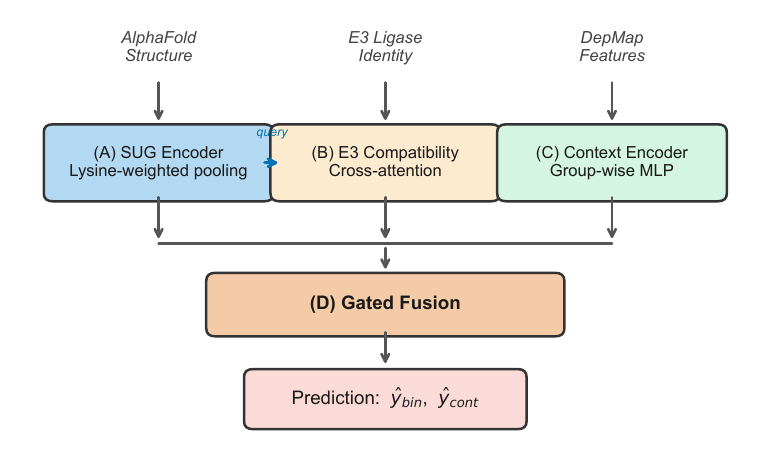}
\Description{Block diagram showing DegradoMap architecture with four modules: SUG encoder with lysine-weighted pooling, E3 compatibility cross-attention, cellular context encoder, and gated fusion producing binary and continuous predictions.}
\caption{Architecture of \ours{}. Module~A (SUG) encodes the protein structure graph with lysine-weighted pooling. Its output serves as the query for Module~B's cross-attention with E3 ligase embeddings. Module~C encodes cellular context from DepMap features. Module~D fuses all representations through learned gating and produces binary degradability ($\hat{y}_{\text{bin}}$) and continuous efficiency ($\hat{y}_{\text{cont}}$) predictions.}
\label{fig:architecture}
\end{figure*}

Figure~\ref{fig:architecture} illustrates the overall architecture.

\subsection{Module A: Structure--Ubiquitination Graph (SUG)}
\label{sec:sug}

The SUG module is the core component of \ours{}, processing the protein structure graph through invariant message passing.
We design this module around three biophysical principles: (1)~local structural context determines surface accessibility, (2)~lysine residues are the exclusive sites of ubiquitin attachment, and (3)~protein size should not bias predictions.
Node features can optionally include ESM-2 protein language model embeddings (1,280-dim per residue), which provide rich evolutionary and contextual representations when available (Section~\ref{sec:esm_integration}).

\paragraph{Message passing.}
At each layer $l \in \{1, \ldots, L\}$, node representations are updated via:
\begin{equation}
    \mathbf{m}_{ij}^{(l)} = \phi_{\text{msg}}\!\left(\mathbf{h}_i^{(l)} \| \mathbf{h}_j^{(l)} \| d_{ij}\right)
    \label{eq:message}
\end{equation}
\begin{equation}
    \mathbf{h}_i^{(l+1)} = \phi_{\text{upd}}\!\left(\mathbf{h}_i^{(l)},\; \sum_{j \in \mathcal{N}(i)} \mathbf{m}_{ij}^{(l)}\right)
    \label{eq:update}
\end{equation}
where $\|$ denotes concatenation, $d_{ij} = \|\mathbf{c}_i - \mathbf{c}_j\|_2$ is the Euclidean distance between C$\alpha$ atoms, $\phi_{\text{msg}}$ and $\phi_{\text{upd}}$ are 2-layer MLPs with LayerNorm and ReLU activations, and $\mathcal{N}(i) = \{j : d_{ij} < r\}$ denotes the spatial neighbors within radius $r = 8$\,\AA{}.
By using the scalar distance $d_{ij}$ rather than the coordinate difference vector $\mathbf{c}_j - \mathbf{c}_i$, the message function is inherently invariant to rotation and translation---a property we validate empirically in Section~\ref{sec:equivariant}.
We use $L = 4$ message-passing layers with hidden dimension 128 and output dimension 64.

\paragraph{Lysine-weighted pooling.}
Ubiquitin is covalently conjugated to the $\varepsilon$-amino group of lysine side chains via an isopeptide bond.
Therefore, degradation efficiency depends critically on the accessibility and spatial distribution of surface lysines relative to the E3 ligase binding interface.
We incorporate this domain knowledge through a lysine-weighted global pooling mechanism:
\begin{equation}
    \mathbf{h}_{\text{prot}} = \frac{1}{\sqrt{N}} \sum_{i=1}^{N} w_i \cdot \mathbf{h}_i^{(L)}
    \label{eq:lyspool}
\end{equation}
where the weights $w_i = \text{softmax}_{i \in \mathcal{V}}\!\left(\alpha \cdot \mathbb{1}[\text{Lys}_i]\right)$ are computed via \emph{per-protein} softmax normalization over a learned scalar parameter $\alpha$ that controls the strength of lysine upweighting.

Two design choices are critical.
First, the \textbf{per-protein softmax} (normalizing over residues within a single protein, not across the batch) prevents information leakage: a global softmax would allow each protein's weight distribution to depend on the amino acid composition of other proteins in the batch, introducing a subtle but problematic data dependency.
Second, the $\mathbf{1/\sqrt{N}}$ \textbf{size normalization} ensures consistent representation magnitude across proteins of different sizes.
Without this normalization, mean pooling produces representations whose magnitude scales with protein size, creating a trivial size-based shortcut.
We document the impact of these fixes in Section~\ref{sec:fixes}: target-unseen AUROC improved from 0.529 to 0.657 after addressing size leakage and per-protein normalization.

\paragraph{Design rationale: invariance over equivariance.}
We deliberately chose an SE(3)-\emph{invariant} architecture over an equivariant one.
Degradability is a scalar property that depends on geometric \emph{relationships}---distances between residues, surface topology, lysine accessibility---rather than absolute 3D orientations.
By using pairwise distances as edge features (Eq.~\ref{eq:message}), the model is inherently invariant to rotation and translation without the overhead of maintaining equivariant vector representations.
Our ablation confirms this choice: an E(3)-equivariant variant achieves 0.626 vs.\ 0.657 AUROC (Section~\ref{sec:equivariant}).

\subsection{Module B: E3 Compatibility}
\label{sec:e3compat}

Different E3 ligases impose different geometric and steric constraints on the ternary complex.
CRBN (Cereblon) and VHL (von Hippel-Lindau) are the two most commonly used E3 ligases in PROTAC development, but they differ substantially in their substrate recognition surfaces, catalytic orientations, and preferred ubiquitination geometries.
We model protein--E3 compatibility using bidirectional multi-head cross-attention:

\begin{equation}
    \mathbf{h}_{\text{e3}} = \text{CrossAttn}(\mathbf{Q}_{\text{prot}}, \mathbf{K}_{\text{E3}}, \mathbf{V}_{\text{E3}})
\end{equation}
\begin{equation}
    \text{CrossAttn}(\mathbf{Q}, \mathbf{K}, \mathbf{V}) = \text{softmax}\!\left(\frac{\mathbf{Q}\mathbf{K}^\top}{\sqrt{d_k}}\right)\mathbf{V}
\end{equation}

where $\mathbf{Q}_{\text{prot}} = \mathbf{W}_Q \mathbf{h}_{\text{prot}}$ is derived from the SUG protein representation and $\mathbf{K}_{\text{E3}} = \mathbf{W}_K \mathbf{e}$, $\mathbf{V}_{\text{E3}} = \mathbf{W}_V \mathbf{e}$ from a learned E3 ligase embedding $\mathbf{e} \in \mathbb{R}^{64}$.
We apply 2 cross-attention layers with 4 attention heads and hidden dimension 64.

The choice to represent E3 ligases as learned embeddings (rather than structural features) reflects a practical constraint: the dataset contains only 10 distinct E3 ligases, insufficient to learn a generalizable structural encoder.
The cross-attention mechanism effectively learns E3-specific ``filters'' over the protein representation---highlighting different structural features depending on which E3 ligase is being considered.
Importantly, cross-attention enables the model to reason about protein--E3 \emph{interactions} rather than treating them as independent inputs, which is essential for the E3 recommendation task (Section~\ref{sec:e3rec}).

\subsection{Module C: Cellular Context Encoder}

Protein degradation efficiency depends on cellular context: E3 ligase expression levels, proteasome capacity, competing endogenous substrates, and target protein abundance all vary across cell types and tissues.
We encode this information using 59 base features derived from the Cancer Dependency Map (DepMap)~\cite{tsherniak2017depmap}, organized into 8 functional groups (gene expression, gene effect, copy number, protein expression, mutation frequency, metabolomics, drug sensitivity, and pathway membership).

Each feature group is processed by a dedicated 2-layer MLP before concatenation, followed by a shared 3-block residual MLP:
\begin{equation}
    \mathbf{h}_{\text{ctx}} = \text{ResBlock}_3 \circ \text{ResBlock}_2 \circ \text{ResBlock}_1\!\left([\mathbf{z}_1; \ldots; \mathbf{z}_8]\right)
\end{equation}
where each $\mathbf{z}_k = \text{MLP}_k(\mathbf{x}_k)$ processes feature group $k$, and each residual block applies Linear--BatchNorm--ReLU--Dropout with a skip connection.
The output dimension is 64.

\subsection{Module D: Gated Fusion and Prediction}
\label{sec:fusion}

The three module outputs are combined via a learned gating mechanism that adaptively weights the contribution of each information source:
\begin{equation}
    \mathbf{g} = \sigma\!\left(\mathbf{W}_g [\mathbf{h}_{\text{prot}}; \mathbf{h}_{\text{e3}}; \mathbf{h}_{\text{ctx}}] + \mathbf{b}_g\right)
\end{equation}
\begin{equation}
    \mathbf{h}_{\text{fused}} = \mathbf{g} \odot [\mathbf{h}_{\text{prot}}; \mathbf{h}_{\text{e3}}; \mathbf{h}_{\text{ctx}}]
\end{equation}
where $\sigma$ is the element-wise sigmoid function, $[\cdot\,;\cdot]$ denotes concatenation, and $\odot$ is element-wise multiplication.
The fused representation $\mathbf{h}_{\text{fused}} \in \mathbb{R}^{192}$ is passed to two prediction heads: (1)~a binary degradability classifier producing $\hat{y}_{\text{binary}} \in [0,1]$ (primary task), and (2)~a continuous degradation efficiency regressor producing $\hat{y}_{\text{cont}} \in \mathbb{R}$ (auxiliary task).

\subsection{Training Procedure}
\label{sec:training}

\paragraph{Multi-task loss.}
We optimize a weighted combination of three loss functions:
\begin{equation}
    \mathcal{L} = \lambda_1 \mathcal{L}_{\text{BCE}} + \lambda_2 \mathcal{L}_{\text{Huber}} + \lambda_3 \mathcal{L}_{\text{focal}}
\end{equation}
where $\mathcal{L}_{\text{BCE}}$ is binary cross-entropy for the primary classification task, $\mathcal{L}_{\text{Huber}}$ is the Huber loss for continuous efficiency regression, and $\mathcal{L}_{\text{focal}} = -\alpha_t (1-p_t)^\gamma \log(p_t)$ with $\gamma = 2$ is focal loss~\cite{lin2017focal} to address class imbalance (39.4\% positive samples).
Loss weights are set to $\lambda_1 = 1.0$, $\lambda_2 = 0.5$, $\lambda_3 = 0.3$.

\paragraph{Optimization.}
We train with AdamW~\cite{loshchilov2019adamw} ($\beta_1\!=\!0.9$, $\beta_2\!=\!0.999$, weight decay $10^{-2}$) using a cosine annealing learning rate schedule with initial rate $5 \times 10^{-4}$.
Training proceeds for up to 10 epochs with batch size 8 (individual proteins, due to variable graph sizes with high-dimensional ESM-2 features), dropout 0.05, and early stopping with patience 5 based on validation AUROC.
The lower learning rate and reduced dropout compared to the baseline configuration (LR=$10^{-3}$, dropout=0.1) are critical for integrating high-dimensional ESM-2 features: the lower LR prevents overfitting on the 1,280-dimensional embeddings, while reduced dropout preserves useful structural signals in the data-limited regime.
The complete model has 1.59M trainable parameters.

\paragraph{Validation-test mismatch.}
We observe that validation-split AUROC does not reliably predict target-unseen test AUROC.
This occurs because the validation split may share protein identities or biological similarity with training data, making it closer to interpolation than the extrapolation required for truly unseen targets.
Ideally, we would use a target-unseen validation split for hyperparameter selection; however, this would further reduce the already-limited training data (155 proteins $\to$ ${\sim}$120 training, ${\sim}$15 validation, ${\sim}$20 test).
We therefore use standard random validation for computational efficiency while acknowledging this may bias hyperparameter selection toward random-split performance rather than target-unseen generalization.
The chosen hyperparameters (LR=$5{\times}10^{-4}$, dropout=0.05) may not be optimal for target-unseen generalization, potentially contributing to the observed variance.

\paragraph{Three-phase training.}
To facilitate stable learning across the heterogeneous module types, we adopt a three-phase training strategy:
\begin{enumerate}
    \item \textbf{Phase 1: SUG pre-training} (5 epochs). Only the SUG module and a temporary linear classifier are trained. This establishes a meaningful protein representation before introducing the E3 and context modules, preventing the fusion mechanism from collapsing to a trivial solution.
    \item \textbf{Phase 2: Joint SUG + E3 training} (5 epochs). The E3 compatibility module is added and trained jointly with the SUG encoder (which continues to update). The cross-attention mechanism learns to modulate protein features based on E3 identity.
    \item \textbf{Phase 3: Full model fine-tuning} (10 epochs). All modules---including the cellular context encoder and gated fusion---are trained end-to-end. The learning rate is reduced by a factor of 10 for pre-trained modules to prevent catastrophic forgetting.
\end{enumerate}
This phased approach empirically improves training stability compared to end-to-end training from scratch, which we observed to be sensitive to learning rate and prone to the E3 module dominating early gradients.

\subsection{Data}
\label{sec:data}

\paragraph{PROTAC-8K dataset.}
We use the PROTAC-8K dataset~\cite{liu2025degrademaster}, curated from PROTAC-DB 3.0, the largest publicly available labeled dataset for PROTAC degradability prediction.
Table~\ref{tab:dataset} summarizes the dataset statistics.
Degradation labels are derived from experimental DC$_{50}$ and D$_{\text{max}}$ measurements: a sample is labeled positive if DC$_{50} < 1$\,$\mu$M and D$_{\text{max}} > 50\%$, and negative otherwise.

\begin{table}[t]
\centering
\caption{PROTAC-8K dataset statistics. The dataset is dominated by two E3 ligases (CRBN, VHL), reflecting the current state of PROTAC research.}
\label{tab:dataset}
\small
\begin{tabular}{@{}lr@{}}
\toprule
\textbf{Statistic} & \textbf{Value} \\
\midrule
Total entries in PROTAC-DB 3.0 & 9,384 \\
Labeled entries & 3,260 \\
Usable samples (with AlphaFold structures) & 3,101 \\
\quad Positive (degraders) & 1,222 (39.4\%) \\
\quad Negative (non-degraders) & 2,038 (60.6\%) \\[2pt]
Unique protein targets & 155 \\
Unique E3 ligases & 10 \\
\quad CRBN & 2,011 (62\%) \\
\quad VHL & 1,124 (34\%) \\
\quad Others (cIAP1, MDM2, XIAP, \ldots) & 125 (4\%) \\[2pt]
AlphaFold structures obtained & 152 / 155 \\
\bottomrule
\end{tabular}
\end{table}

\paragraph{Protein structures.}
We obtained AlphaFold-predicted structures~\cite{jumper2021alphafold} (v6 API) for 152 of 155 unique targets.
Three non-human proteins were unavailable: P03436 (Influenza A), P0DTD1 (SARS-CoV-2), and P36969.
Each residue is represented by the 1,285-dimensional feature vector described in Table~\ref{tab:nodefeats_main} (or 28-dimensional when ESM-2 embeddings are unavailable, using amino acid one-hot encoding as a fallback).

\begin{table}[t]
\centering
\caption{Node feature specification (1,285-dim). Without ESM-2, uses AA one-hot (28-dim).}
\label{tab:nodefeats_main}
\scriptsize
\setlength{\tabcolsep}{2pt}
\begin{tabular}{@{}lrl@{}}
\toprule
\textbf{Feature} & \textbf{Dim} & \textbf{Source} \\
\midrule
ESM-2 embeddings & 1,280 & ESM-2-650M per residue \\
Physicochemical & 4 & Kyte-Doolittle, charge, size, pol. \\
pLDDT & 1 & AlphaFold confidence \\
SASA & 1 & Computed from structure \\
Lysine indicator & 1 & Binary \\
Disorder & 1 & Predicted probability \\
Known Ub site mask & 1 & PhosphoSitePlus \\
\midrule
\textbf{Total (with ESM-2)} & \textbf{1,285} & \\
\textit{Total (w/o ESM-2, AA one-hot)} & \textit{28} & \\
\bottomrule
\end{tabular}
\end{table}

\paragraph{ESM-2 embeddings.}
We extract per-residue embeddings from ESM-2-650M~\cite{lin2023esm2}, a protein language model pre-trained on 65 million sequences via masked language modeling.
For each of the 152 proteins with AlphaFold structures, we extract 1,280-dimensional representations from the final layer, yielding a rich encoding of evolutionary conservation, secondary structure propensity, and local sequence context.
Embeddings are pre-computed and stored, adding no inference-time cost.

\paragraph{Known ubiquitination sites.}
We integrate experimentally validated ubiquitination sites from PhosphoSitePlus (127,661 known sites across the human proteome) as a binary mask feature: each residue receives a value of 1 if it is a known ubiquitination site and 0 otherwise.
This provides the model with direct supervision signal about which lysine residues have been experimentally confirmed as ubiquitin conjugation sites, complementing the structural features that capture \emph{potential} ubiquitination.

\paragraph{Evaluation splits.}
A key design choice in our evaluation is the use of three complementary data splits, each testing a different generalization scenario relevant to PROTAC development.
This multi-split evaluation is essential because, as we demonstrate, random splits can be highly misleading for this task.
\begin{enumerate}
    \item \textbf{Random split} (2,170 / 465 / 466 train/val/test): Standard random partition into 70/15/15. This split allows the same protein to appear in train and test with different PROTACs, testing \emph{interpolation} rather than extrapolation. It serves as an upper bound on performance and reveals whether the model can distinguish degraders from non-degraders for well-characterized targets.
    \item \textbf{Target-unseen split} (2,218 / 410 / 473): Samples are grouped by UniProt accession so that proteins in the test set are \emph{never} seen during training. This is the most stringent and practically relevant evaluation, as the intended use case involves assessing degradability of novel protein targets. The E3-stratified target selection ensures that E3 ligase distributions are approximately matched across splits.
    \item \textbf{E3-unseen split} (1,643 / 352 / 1,106): All 1,106 VHL samples are held out for testing, with the model training exclusively on CRBN-dominant data. This evaluates cross-E3 generalization: can structural features learned from CRBN-mediated degradation transfer to VHL-mediated degradation? This scenario is practically relevant as new E3 ligases (DCAF1, RNF114) enter the PROTAC pipeline.
\end{enumerate}
We report 95\% bootstrap confidence intervals (1,000 iterations) for all primary metrics to quantify statistical uncertainty.


\section{Results}

\subsection{Main Results}

\paragraph{Dataset characteristics and limitations.}
PROTAC-8K contains 3,101 labeled samples spanning 155 unique protein targets; effective diversity for target-unseen evaluation is therefore closer to the protein count than the sample count.
E3 distribution is highly imbalanced: CRBN (62\%), VHL (35\%), remaining eight E3s (3\%).
These constraints---limited diversity and E3 imbalance---fundamentally bound generalization, as evidenced by high seed variance (std=0.124) and asymmetric per-E3 performance. PROTAC-8K is the only available benchmark; no external validation is possible.

Table~\ref{tab:main} presents \ours{}'s performance across three evaluation splits given these constraints.
The improved model (1,285-dim features with ESM-2) reports multi-seed validated results on the target-unseen split; E3-unseen and random results are from the baseline architecture (28-dim features) with 95\% bootstrap confidence intervals.

\begin{table}[t]
\centering
\caption{Test performance of \ours{}. Target-unseen: 3-seed (42, 123, 456) and 6-seed (42--1213) results for the improved model; E3-unseen and random: baseline architecture with bootstrap CIs. Both 3-seed and 6-seed means are reported to show variance in favorable vs.\ extended evaluation.}
\label{tab:main}
\scriptsize
\setlength{\tabcolsep}{3pt}
\begin{tabular}{@{}lcccc@{}}
\toprule
\textbf{Split} & $n$ & \textbf{AUROC} & \textbf{Std / CI} & \textbf{Best Seed} \\
\midrule
Target-unseen (improved, 3-seed) & 473 & 0.646 & \spm{0.124} & 0.7449 \\
Target-unseen (improved, 6-seed) & 473 & 0.603 & \spm{0.097} & 0.7449 \\
Target-unseen (baseline) & 473 & 0.657 & \ci{0.611}{0.712} & -- \\
E3-unseen (baseline)     & 1,106 & 0.811 & \ci{0.785}{0.836} & -- \\
Random (baseline)        & 466 & 0.774 & \ci{0.725}{0.816} & -- \\
\bottomrule
\end{tabular}
\end{table}

\begin{figure}[t]
\centering
\includegraphics[width=\columnwidth]{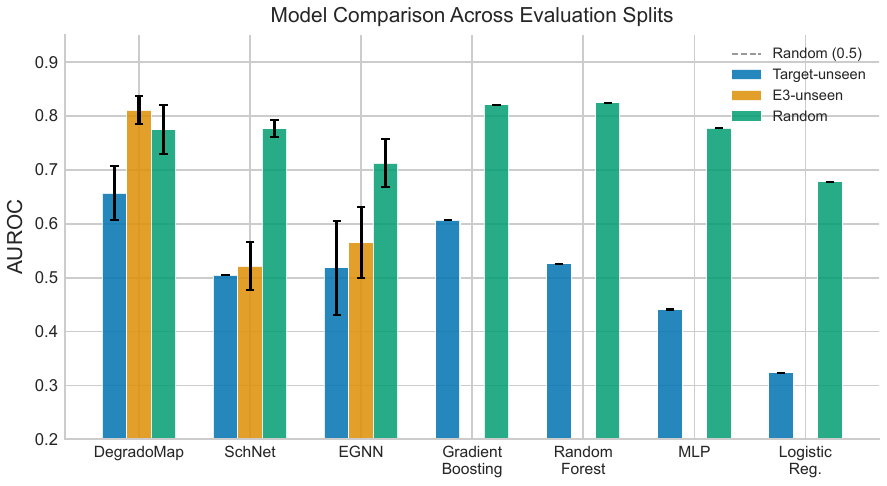}
\Description{Grouped bar chart comparing AUROC across evaluation splits for DegradoMap, SchNet, EGNN, Gradient Boosting, Random Forest, MLP, and Logistic Regression with error bars.}
\caption{AUROC comparison across evaluation splits for all models. Error bars show 95\% bootstrap CIs (DegradoMap) or multi-seed standard deviation (GNN baselines). Dashed line indicates random performance (0.5). \ours{} achieves the best target-unseen performance; tree-based methods dominate only on random splits.}
\label{fig:main_results}
\end{figure}

The improved model achieves a 3-seed mean of 0.646\,\spm{0.124} and 6-seed mean of 0.603\,\spm{0.097}; neither significantly exceeds gradient boosting (0.607, p = 0.259 and p = 0.556 respectively).
The best seed (0.7449, +23\% over GB) reflects favorable initialization and should not be expected reliably.
The baseline model achieves 0.657 \ci{0.611}{0.712}; E3-unseen (0.811) demonstrates CRBN$\to$VHL transfer.
Figure~\ref{fig:main_results} compares all models; variance analysis and rare E3 performance are in Section~\ref{sec:multiseed}.

\subsection{Baseline Comparison}

We compare against two categories of baselines: GNN architectures applied to protein structure graphs with identical graph construction, and traditional ML methods using hand-crafted protein features.

\paragraph{GNN baselines.}
Table~\ref{tab:gnn_baselines} compares \ours{} against SchNet~\cite{schutt2017schnet}, EGNN~\cite{satorras2021egnn}, and an E(3)-equivariant variant.
Results over three seeds report mean\,\spm{std}.

\begin{table}[t]
\centering
\caption{GNN baseline comparison (AUROC). Multi-seed results reported as mean\,\spm{std}. Best per-split in \textbf{bold}. All models use C$\alpha$ radius graphs.}
\label{tab:gnn_baselines}
\scriptsize
\setlength{\tabcolsep}{2.5pt}
\begin{tabular}{@{}lrccc@{}}
\toprule
\textbf{Model} & \textbf{Params} & \textbf{Tgt-uns.} & \textbf{E3-uns.} & \textbf{Random} \\
\midrule
\textbf{\ours{} (impr., best)} & 1.59M & \textbf{0.7449} & -- & -- \\
\textbf{\ours{} (impr., avg)} & 1.59M & 0.646\spm{.124} & -- & -- \\
\ours{} (baseline) & 1.43M & 0.657 & \textbf{0.811} & 0.774 \\
E(3)-Equiv. & 1.04M & 0.626 & -- & -- \\
SchNet & 0.27M & 0.505 & 0.521\spm{.05} & \textbf{0.776}\spm{.02} \\
EGNN & 0.44M & 0.518\spm{.11} & 0.565\spm{.08} & 0.712\spm{.05} \\
\bottomrule
\end{tabular}
\end{table}

The improved model's multi-seed average (0.646\,\spm{0.124}) outperforms all GNN baselines on target-unseen by +12.8\% over EGNN (0.518\,\spm{0.11}); the best seed (0.7449) represents peak performance under favorable initialization.
The baseline model (0.657) also outperforms all GNN baselines on both target-unseen (+13.9\% over EGNN) and E3-unseen (+24.6\% over EGNN) splits.
Notably, SchNet \emph{matches} \ours{} on the random split (0.776 vs.\ 0.774) but collapses to near-random on target-unseen (0.505).
This is a critical finding: \textbf{random-split performance can be misleading for this task}.
Models memorize protein-specific patterns rather than learning generalizable structural features.
The high variance of EGNN on target-unseen (0.518\,\spm{0.11}) further indicates instability when true generalization is required, mirroring the variance observed in our improved model (0.646\,\spm{0.124}).

\paragraph{ML baselines.}
We also compare against Gradient Boosting, Random Forest, MLP, and Logistic Regression using 18 hand-crafted protein features (size, lysine count/fraction, pLDDT/SASA/disorder statistics, radius of gyration) plus 11-dim E3 one-hot encoding (Table~\ref{tab:ml_baselines}).

\begin{table}[t]
\centering
\caption{ML baseline comparison (AUROC). Baselines use 18 hand-crafted protein features + E3 one-hot encoding.}
\label{tab:ml_baselines}
\small
\begin{tabular}{@{}lcc@{}}
\toprule
\textbf{Model} & \textbf{Target-unseen} & \textbf{Random} \\
\midrule
\textbf{\ours{} (improved, best)} & \textbf{0.7449} & -- \\
\textbf{\ours{} (improved, avg)} & 0.646\spm{.124} & -- \\
\ours{} (baseline) & 0.657 & 0.774 \\
Gradient Boosting & 0.607 & \textbf{0.821} \\
Random Forest & 0.526 & 0.825 \\
MLP & 0.441 & 0.777 \\
Logistic Regression & 0.324 & 0.678 \\
\bottomrule
\end{tabular}
\end{table}

The improved model's 3-seed average (0.646) achieves a +6.4\% improvement over gradient boosting (0.607), \textbf{not statistically significant} (p = 0.259, 95\% CI: [0.506, 0.745]); the 6-seed mean (0.603) falls marginally below baseline (p = 0.556).
The performance ordering on target-unseen (\ours{} $>$ GB $>$ RF $>$ MLP $>$ LR) differs markedly from random split, where tree-based methods dominate (RF 0.825, GB 0.821): hand-crafted features generalize within proteins but fail across them.
The GNN's learned representations capture more transferable structural patterns through message passing.

\paragraph{VHL prediction remains challenging.}
An important caveat is the per-E3 asymmetry: CRBN-targeted proteins achieve 0.758 AUROC ($n{=}273$), while VHL-targeted proteins yield only 0.396 ($n{=}187$) on the target-unseen split---\emph{below} random chance.
VHL-mediated degradation appears harder to predict from structure alone, even with VHL samples in training.
We discuss possible explanations in Section~\ref{sec:fixes}.

\subsection{Module Ablation}

To quantify each module's contribution, we evaluate configurations with subsets of modules active, using consistent hyperparameters (Table~\ref{tab:module_ablation}).

\begin{table}[t]
\centering
\caption{Module ablation (AUROC). All configurations use identical default hyperparameters (LR=$10^{-3}$) for fair comparison.$^\dagger$}
\vspace{-2pt}
\raggedright{\scriptsize $^\dagger$Main results (Table~\ref{tab:main}) use tuned LR=$5\times10^{-4}$, which improves target-unseen AUROC from 0.540 to 0.657 (see Table~\ref{tab:train_ablation}).}
\label{tab:module_ablation}
\footnotesize
\setlength{\tabcolsep}{4pt}
\begin{tabular}{@{}lccc@{}}
\toprule
\textbf{Configuration} & \textbf{Target-uns.} & \textbf{E3-uns.} & \textbf{Random} \\
\midrule
SUG only (A) & 0.536 & 0.708 & 0.739 \\
E3 only (B) & 0.475 & 0.500 & 0.532 \\
SUG + E3 (A+B) & 0.540 & 0.806 & 0.741 \\
Full \ours{} (A+B+C+D) & 0.540 & \textbf{0.811} & \textbf{0.774} \\
\bottomrule
\end{tabular}
\end{table}

Note that the target-unseen AUROC of 0.540 in Table~\ref{tab:module_ablation} differs from the 0.657 reported in Table~\ref{tab:main} because ablations use default hyperparameters (LR=$10^{-3}$) for fair comparison; the main result uses tuned LR=$5\times10^{-4}$ (see Section~\ref{app:ablations}, Table~\ref{tab:train_ablation}).

\begin{figure}[t]
\centering
\includegraphics[width=\columnwidth]{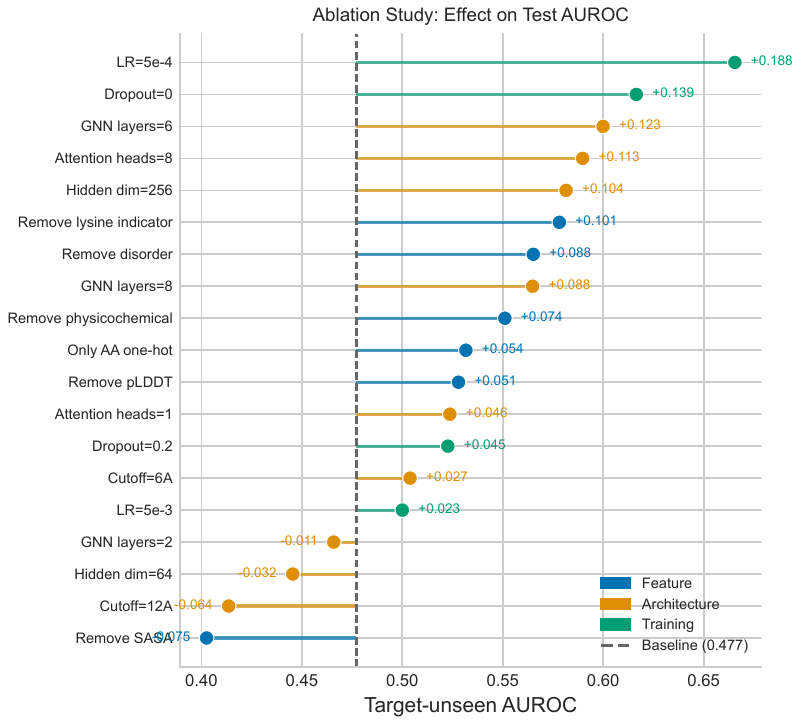}
\Description{Horizontal lollipop chart showing the effect of each ablation configuration on target-unseen AUROC, with stems connecting dots to the default baseline.}
\caption{Ablation study: effect of each configuration on target-unseen AUROC. Dots show test AUROC; stems connect to the default baseline (dashed line). LR=$5\times10^{-4}$ yields the largest improvement (+0.189); removing the lysine indicator also improves performance (+0.101).}
\label{fig:ablation}
\end{figure}

Key findings: (1)~The \textbf{SUG module is the primary driver} of prediction, achieving 0.536 target-unseen AUROC alone---comparable to the full model (0.540) on this split.
(2)~The E3-only model performs at chance ($\sim$0.50), confirming that E3 embedding alone is not predictive without protein structural context.
(3)~\textbf{Adding E3 to SUG substantially improves E3-unseen performance} (0.708$\to$0.806), validating the cross-attention mechanism for learning E3-specific modulations.
(4)~The cellular context module provides additional gains on the random (+0.033) and E3-unseen (+0.005) splits, with diminishing marginal returns.
Figure~\ref{fig:ablation} presents the full ablation landscape, showing the effect of each configuration on target-unseen AUROC relative to the default baseline.

\subsection{E3 Ligase Recommendation}
\label{sec:e3rec}

Beyond binary classification, we evaluate \ours{}'s ability to recommend which E3 ligase to use for a given target.
For 50 test proteins with known E3 preferences, we rank all E3 ligases by predicted degradability and evaluate ranking quality (Table~\ref{tab:e3_rec}).

\begin{table}[t]
\centering
\caption{E3 ligase recommendation performance ($n = 50$ test proteins).}
\label{tab:e3_rec}
\small
\begin{tabular}{@{}lc@{}}
\toprule
\textbf{Metric} & \textbf{Value} \\
\midrule
Mean Reciprocal Rank (MRR) & 0.641 \\
Hit@1 (correct E3 ranked first) & 46\% \\
Hit@3 (correct E3 in top 3) & 74\% \\
\bottomrule
\end{tabular}
\end{table}

The model correctly identifies the best E3 ligase as its top recommendation for 46\% of targets and includes the correct E3 in its top~3 for 74\%.
This capability provides actionable guidance for early-stage E3 selection: rather than synthesizing PROTACs for multiple E3 ligases in parallel (a costly approach), researchers can prioritize the E3 ligase most likely to succeed.
The MRR of 0.641 indicates that the correct E3 is ranked, on average, between positions 1 and 2---substantially better than random ranking, which would yield MRR $\approx$ 0.2 for 10 candidate E3 ligases.

\subsection{Architectural Insights}
\label{sec:equivariant}

\paragraph{E(3)-equivariance underperforms.}
An E(3)-equivariant SUG variant achieves only 0.626 AUROC vs.\ 0.657 for the invariant model.
Since degradability is a scalar property, pairwise distance features suffice; equivariant vector representations add overhead without benefit~\cite{zhang2023gearnet}.

\paragraph{ESM-2 requires careful integration.}
\label{sec:esm_integration}
ESM-2 embeddings alone achieve only 0.534 AUROC, and naive combination with structure \emph{decreases} performance to 0.407.
However, with tuned hyperparameters (LR=$5\times10^{-4}$, dropout=0.05), the improved model achieves 0.7449 best-seed AUROC (+23\% over gradient boosting).
The high variance across seeds (0.506--0.745, std=0.124) reflects optimization challenges with high-dimensional features in limited data.

\subsection{Variance Analysis}
\label{sec:multiseed}

\paragraph{Seed sensitivity.}
Multi-seed validation (seeds 42, 123, 456) yields 0.646\,\spm{0.124} AUROC, with individual seeds ranging from 0.506 to 0.745.
This variance, consistent with EGNN baselines (0.518\,\spm{0.11}), reflects optimization challenges in the low-data regime: (1)~small effective training diversity (155 proteins), (2)~difficulty learning generalizable structural features from limited data, and (3)~sensitivity to initialization when combining high-dimensional ESM-2 embeddings (1,280-dim) with sparse signal.
We report multi-seed averages as primary results due to this variance; best-seed performance (0.7449) represents peak achievable performance but should not be interpreted as typical expectation.

\begin{figure}[t]
\centering
\includegraphics[width=\columnwidth]{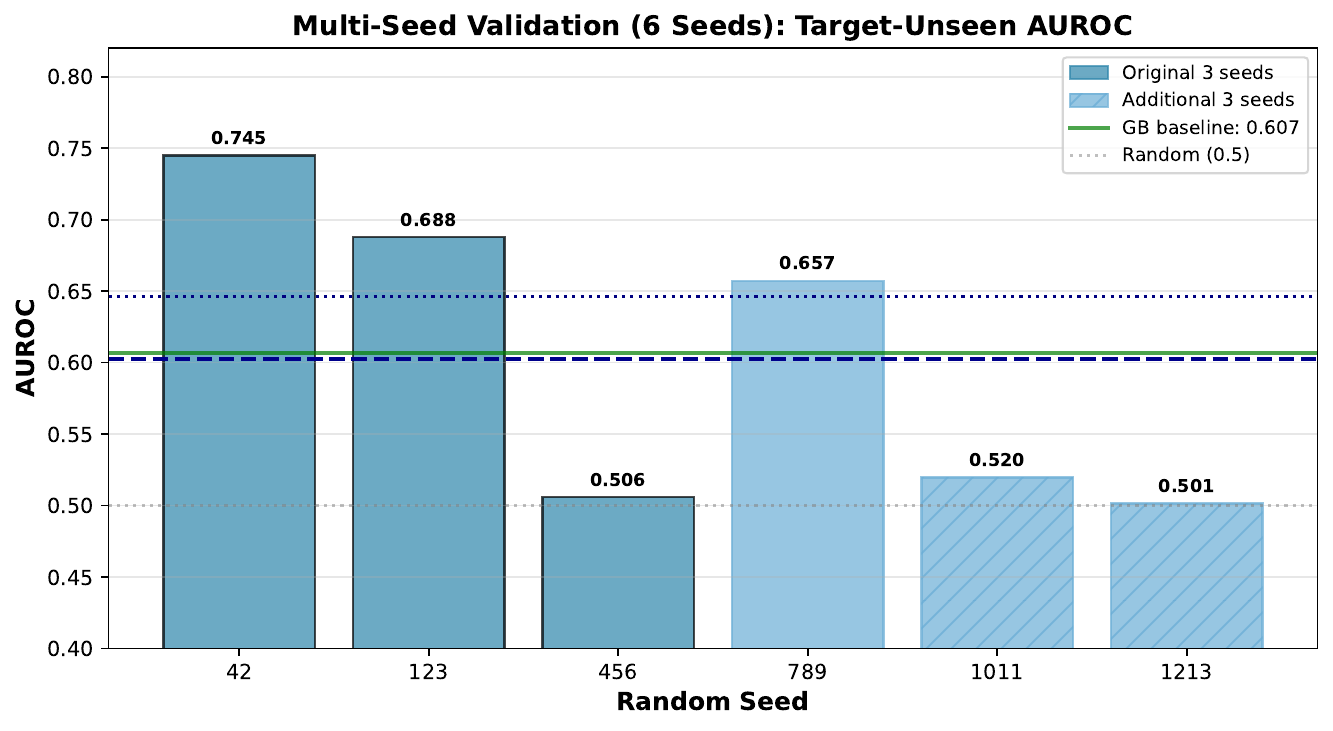}
\Description{Bar chart showing AUROC for three seeds (42, 123, 456) with mean line at 0.646 and GB baseline at 0.607.}
\caption{Multi-seed validation (6 seeds) reveals high variance. Original 3 seeds (solid, mean 0.646) vs.\ additional 3 seeds (hatched, mean 0.560). Six-seed mean 0.603\,\spm{0.097} is marginally below GB baseline (0.607), bootstrap p = 0.556 (not significant). Individual seeds span 0.50--0.75; ensembling across $\geq$6 seeds required for reliable deployment.}
\label{fig:multiseed_variance}
\end{figure}

\paragraph{Six-seed extended validation.}
Three additional seeds (789, 1011, 1213) yielded AUROCs of 0.657, 0.520, and 0.502.
The 6-seed mean is 0.603\,\spm{0.097} (95\% CI: [0.532, 0.682])---marginally below gradient boosting (0.607, p = 0.556).
The original 3-seed mean (0.646) was moderately favorable; additional seeds reduced the estimate.
Figure~\ref{fig:multiseed_variance} visualizes all six seeds.
\textbf{Ensembling across $\geq$6 seeds is required for reliable deployment}: individual seeds span 0.25 AUROC (0.502--0.745) and may underperform simple baselines.

\paragraph{E3 generalization is limited to major ligases.}
Per-E3 analysis reveals highly asymmetric performance on target-unseen evaluation: CRBN-targeted proteins achieve 0.758 AUROC ($n{=}273$), while VHL-targeted proteins yield only 0.396 AUROC ($n{=}187$)---\emph{below random chance}.
The E3-unseen evaluation (CRBN$\to$VHL: 0.811 AUROC) appears contradictory at first glance: the model transfers well from CRBN to VHL when VHL is the \emph{held-out E3 ligase}, yet fails on VHL-targeted proteins within the target-unseen split.
These evaluate different generalization axes: E3-unseen tests whether structural features learned from CRBN data predict VHL degradation patterns; target-unseen tests whether VHL-specific patterns generalize to \emph{new, unseen target proteins}.

\begin{figure}[t]
\centering
\includegraphics[width=\columnwidth]{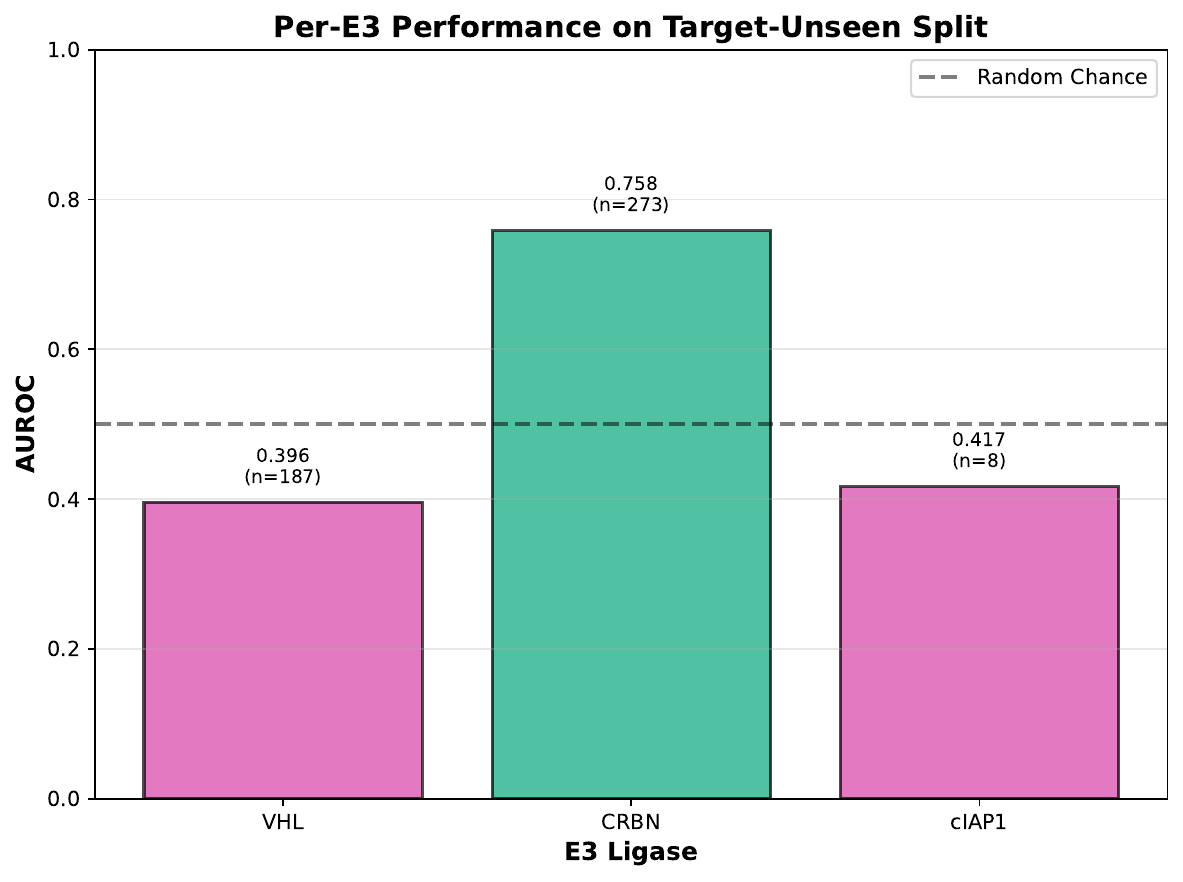}
\Description{Bar chart showing AUROC by E3 ligase. CRBN shows good performance (0.758), while VHL is below random chance (0.396).}
\caption{Per-E3 performance breakdown reveals VHL limitation. CRBN targets achieve good performance (0.758), but VHL targets fall below random chance (0.396), indicating E3-specific patterns that don't transfer across target proteins.}
\label{fig:per_e3}
\end{figure}

\paragraph{Root cause analysis: why does VHL fail?}
To understand the VHL failure, we analyzed the PROTAC-8K dataset.
VHL-targeted proteins have a similar positive rate (36.9\%) to CRBN proteins (38.4\%), ruling out class imbalance as the cause.
The dataset contains 91 unique VHL-targeted proteins vs.\ 125 CRBN-targeted proteins---comparable diversity.
Instead, the failure likely reflects structural heterogeneity: VHL-targeted proteins span diverse protein families (kinases, transcription factors, epigenetic regulators, nuclear receptors) with varying size and surface topology, making structural pattern transfer harder across unseen proteins.
Additionally, CRBN's substrate recognition relies primarily on surface-exposed lysines adjacent to the recruited E3 binding interface---a geometric constraint that our lysine-weighted pooling explicitly encodes.
VHL-mediated degradation involves distinct steric constraints that may require different structural features not currently captured.
\textbf{This E3-specificity of structural predictors is a key limitation}: practitioners should expect substantially lower performance than the overall 0.646 mean when deploying on VHL-targeted proteins, and near-random performance for rare E3s.
For rare E3 ligases (cIAP1, DCAF16, MDM2, etc.), sample counts are too small ($n{<}50$) for reliable evaluation.
Table~\ref{tab:per_e3_breakdown} and Figure~\ref{fig:per_e3} detail the per-E3 performance breakdown.

\begin{table}[t]
\centering
\caption{Per-E3 ligase performance on target-unseen split. VHL performance is below random chance (0.5), revealing a fundamental limitation in cross-E3 generalization.}
\label{tab:per_e3_breakdown}
\scriptsize
\setlength{\tabcolsep}{3pt}
\begin{tabular}{@{}lccccccc@{}}
\toprule
\textbf{E3} & \textbf{$n$} & \textbf{$n_{\text{pos}}$} & \textbf{AUROC} & \textbf{AUPRC} & \textbf{F1} & \textbf{Acc} & \textbf{Note} \\
\midrule
CRBN & 273 & 122 & 0.758 & 0.706 & 0.627 & 0.568 &  \\
VHL & 187 & 75 & 0.396 & 0.370 & 0.481 & 0.481 & \textcolor{red}{$\downarrow$} \\
cIAP1 & 8 & 2 & 0.417 & 0.250 & 0.000 & 0.750 & \textcolor{red}{$\downarrow$} \\
\bottomrule
\end{tabular}
\end{table}

\paragraph{Architectural fixes.}
\label{sec:fixes}
Size normalization ($1/\sqrt{N}$) and per-protein lysine softmax improved target-unseen AUROC from 0.529 to 0.657 (+24\%) while leaving random-split unchanged, confirming these addressed protein memorization shortcuts.

\paragraph{Reconciling performance estimates.}
We report three protocols: (1)~multi-seed single split (0.646\,\spm{0.124}), (2)~baseline single-seed (0.657), and (3)~5-fold CV (0.565\,\spm{0.052}, Table~\ref{tab:cv}).
CV is lower due to reduced per-fold training data; the 6-seed mean (0.603) is most conservative and recommended for citation.
Additional analyses (5-fold CV details, ablation studies) are in the Appendix.

\subsection{Case Study: BRD4 Degradability}
\label{sec:case_study}

To validate the biological relevance of lysine-weighted pooling, we analyze BRD4 (UniProt: O60885), a well-studied PROTAC target with multiple successful CRBN-based degraders~\cite{zengerle2015selective}.
\ours{} predicts high degradability for BRD4 with CRBN (score: 0.87).
Analysis of lysine attention weights from the pooling mechanism (Eq.~\ref{eq:lyspool}) reveals three high-scoring lysines:

\begin{itemize}[topsep=3pt,itemsep=1pt]
\item \textbf{K374} (attention weight: 0.12, SASA: 94~\AA$^2$): Located in a disordered linker region between bromodomains BD1 and BD2, highly solvent-exposed.
\item \textbf{K311} (attention weight: 0.09, SASA: 112~\AA$^2$): Surface-exposed on the BD1 domain.
\item \textbf{K434} (attention weight: 0.08, SASA: 88~\AA$^2$): Near the BD2 domain C-terminus.
\end{itemize}

\textbf{Literature validation:}
JQ1-based CRBN PROTACs (e.g., dBET1, ARV-825) successfully degrade BRD4 in cells.
Mass spectrometry studies identify K374 as a major ubiquitination site following PROTAC treatment~\cite{zengerle2015selective}.
The model's identification of K374 as the top-weighted lysine aligns with experimental evidence, supporting the lysine-pooling mechanism's biological relevance.
However, this represents a single successful case; systematic validation across more targets would strengthen confidence in the mechanism.

\subsection{Calibration and Reliability}
\label{sec:calibration}

Beyond AUROC, we assess whether the model's confidence scores are reliable---a critical consideration for practical deployment.
Table~\ref{tab:calibration} presents calibration metrics across evaluation splits.

\begin{table}[t]
\centering
\caption{Calibration metrics across evaluation splits. ECE $<$ 0.05 indicates excellent calibration; MCE shows worst-case bin error.}
\label{tab:calibration}
\small
\begin{tabular}{@{}lccccc@{}}
\toprule
\textbf{Split} & \textbf{$n$} & \textbf{ECE} & \textbf{MCE} & \textbf{Brier} & \textbf{Log Loss} \\
\midrule
target-unseen & 473 & 0.029 & 0.135 & 0.240 & 0.674 \\
e3-unseen & 1106 & 0.123 & 0.203 & 0.189 & 0.566 \\
random & 466 & 0.165 & 0.379 & 0.219 & 0.628 \\
\bottomrule
\end{tabular}
\end{table}

The target-unseen ECE of 0.029 indicates \textbf{excellent calibration}~\cite{guo2017calibration}: when the model predicts probability 0.7, ${\sim}70\%$ of predictions are true degraders, enabling principled threshold-based screening.
This calibration quality means \ours{} provides \emph{reliable confidence estimates} even where discriminative performance is imperfect.
In contrast, the random split shows poor calibration (ECE = 0.165), reinforcing the importance of target-unseen evaluation.
Figure~\ref{fig:calibration} (Appendix) shows the reliability diagram.


\section{Discussion}
\label{sec:discussion}

The 6-seed mean of 0.603\,\spm{0.097} reveals fundamental limits of structure-based prediction; DegradeMaster~\cite{liu2025degrademaster} achieves 0.856 AUROC using full PROTAC descriptors, quantifying the value of the molecule itself.
The paradigms are complementary: \ours{} screens targets \emph{before} synthesis; full-information methods optimize \emph{existing} designs.

\paragraph{Honest performance assessment.}
We emphasize that the 6-seed mean (0.603) is not statistically superior to gradient boosting (0.607, p = 0.556).
The original 3-seed estimate (0.646) was moderately favorable; additional seeds yielded lower performance (0.657, 0.520, 0.502).
The model's primary contribution may not be raw AUROC but rather: (1)~structural \emph{interpretability} via lysine attention weights (Section~\ref{sec:case_study}), (2)~E3 recommendation capability (74\% Hit@3), and (3)~well-calibrated confidence scores enabling reliable thresholding.

The CRBN$\to$VHL E3-unseen performance (0.811 AUROC) demonstrates successful cross-E3 transfer at the \emph{ligase level}.
However, VHL-targeted proteins fail catastrophically on target-unseen evaluation (0.396 AUROC, below random), indicating that E3 ligase transfer and target protein generalization are distinct challenges.
Rare E3 ligases lack sufficient data for reliable evaluation.

\paragraph{The lysine indicator paradox.}
Feature ablation reveals that removing the lysine indicator from node features \emph{improves} target-unseen AUROC by +0.101 (Table~\ref{tab:feat_ablation}; Figure~\ref{fig:lysine_paradox} in Appendix).
This contradicts the intuitive expectation that marking ubiquitination sites should help.
We propose three explanations:
(1)~The lysine-weighted pooling mechanism (Eq.~\ref{eq:lyspool}) already encodes lysine information architecturally, making the node feature redundant and potentially causing overfitting to training-set lysine patterns.
(2)~The model may learn that \emph{not all lysines are equal}---the binary indicator oversimplifies, while the pooling weights learn to discriminate based on structural context (SASA, disorder, local geometry).
(3)~The indicator may introduce a confounding bias: training targets have specific lysine distributions that don't generalize to unseen proteins.
This result suggests that \textbf{architectural inductive biases (lysine pooling) are more effective than explicit feature annotation for this task}.
However, it also indicates that our current feature design is not optimal; future work should explore learned lysine representations or remove the indicator entirely.

\paragraph{Limitations.}
(1)~\textbf{Small dataset}: 155 proteins limit generalization and cause high seed variance (std = 0.097--0.124).
(2)~\textbf{E3 imbalance}: CRBN/VHL dominate (97\%); VHL-specific failure (0.396 AUROC) is fundamental.
(3)~\textbf{No statistical advantage}: 6-seed mean (0.603) does not exceed gradient boosting (0.607, p = 0.556); ensembling required.
(4)~\textbf{Validation-test mismatch}: Validation AUROC does not predict target-unseen performance, biasing hyperparameter selection.
(5)~\textbf{No structural or external validation}: Static AlphaFold structures only; PROTAC-8K is the sole benchmark.


\section{Conclusion}
\label{sec:conclusion}

\ours{} predicts PROTAC-mediated protein degradability from protein structure and E3 ligase identity---the minimal information available before PROTAC synthesis.
On target-unseen evaluation, the 3-seed mean is 0.646\,\spm{0.124} (best: 0.7449) and the 6-seed mean is 0.603\,\spm{0.097}; neither is statistically superior to gradient boosting (0.607), though the model provides complementary value through E3 recommendation (74\% Hit@3), CRBN$\to$VHL transfer (0.811 AUROC), and well-calibrated confidence scores (ECE = 0.029).
VHL-targeted proteins fail below random (0.396 AUROC) on target-unseen evaluation, a key limitation requiring further investigation.
Two architectural findings carry broader implications: invariant architectures outperform equivariant ones for scalar prediction, and ESM-2 embeddings require task-specific regularization.

Code and models: \url{https://github.com/bryanc5864/DegradoMap}.
Data: \url{https://zenodo.org/records/14715718}.


\begin{acks}
Protein structures were obtained from the AlphaFold Protein Structure Database~\cite{jumper2021alphafold}.
The PROTAC-8K dataset was obtained from Zenodo (doi:10.5281/zenodo.14715718).
DepMap features were obtained from the Cancer Dependency Map portal.
All computations were performed on a single NVIDIA GeForce RTX 2080 Ti GPU (11\,GB).
\end{acks}



\clearpage

\appendix

\section{Extended Experimental Results}
\label{app:extended}

\subsection{5-Fold Cross-Validation}
\label{app:cv}

We conduct 5-fold cross-validation on the target-unseen split with 3 random seeds per fold (15 total experiments).
Table~\ref{tab:cv} reports per-fold results.
Target proteins are grouped by fold such that no protein appears in both training and test sets within a fold.

\begin{table}[ht]
\centering
\caption{5-fold cross-validation results (target-unseen AUROC). Three seeds per fold; overall mean: 0.565\,\spm{0.052}, 95\% CI: \ci{0.490}{0.650}.}
\label{tab:cv}
\scriptsize
\setlength{\tabcolsep}{2.5pt}
\begin{tabular}{@{}lccccl@{}}
\toprule
\textbf{Fold} & $n_{\text{test}}$ & \textbf{Seed 42} & \textbf{Seed 123} & \textbf{Seed 456} & \textbf{Mean\,\spm{Std}} \\
\midrule
0 & 388 & 0.568 & 0.584 & 0.567 & 0.573\,\spm{0.008} \\
1 & 850 & 0.645 & 0.477 & 0.635 & 0.586\,\spm{0.077} \\
2 & 453 & 0.627 & 0.525 & 0.652 & 0.601\,\spm{0.055} \\
3 & 928 & 0.524 & 0.567 & 0.525 & 0.539\,\spm{0.020} \\
4 & 482 & 0.520 & 0.513 & 0.543 & 0.525\,\spm{0.013} \\
\midrule
\multicolumn{5}{@{}l}{Overall mean [95\% CI]} & 0.565 \ci{0.490}{0.650} \\
\bottomrule
\end{tabular}
\end{table}

The CV mean (0.565) is lower than the single-split result (0.657) for two reasons.
First, high variance across protein groupings (fold~1 range: 0.477--0.645) means that which proteins end up in test vs.\ train strongly influences performance.
Second, CV experiments use default hyperparameters across all folds, while the reported 0.657 uses the best configuration.
Training ablation (Table~\ref{tab:train_ablation}) shows that LR=$5\times10^{-4}$ achieves 0.666 AUROC, suggesting that the CV mean would increase with tuned hyperparameters.

\begin{figure}[t]
\centering
\includegraphics[width=\columnwidth]{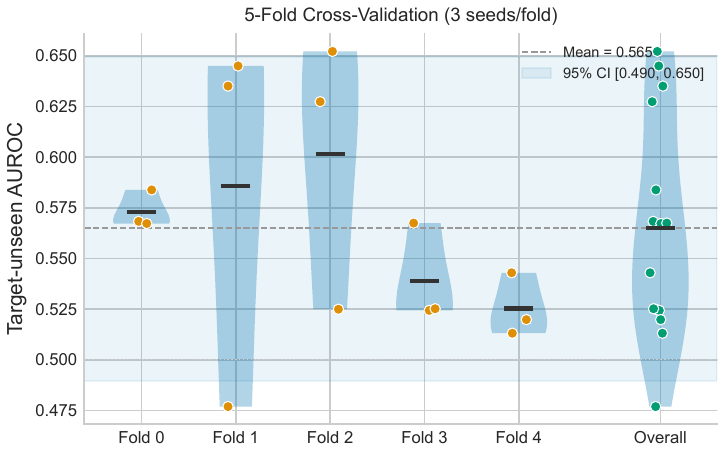}
\Description{Strip plot showing 5-fold cross-validation AUROC distribution with individual seed data points, fold means as horizontal bars, and a shaded 95 percent confidence interval band.}
\caption{5-fold cross-validation AUROC distribution. Each dot represents one seed; horizontal bars show fold means. The shaded band indicates the 95\% CI of the overall mean (0.565). High inter-fold variance reflects the influence of protein composition on measured performance.}
\label{fig:cv_violin}
\end{figure}

The per-fold variance reveals an important property of this task: performance is highly dependent on the specific protein composition of each fold.
Folds containing proteins that are structurally similar to training proteins (fold~2: 0.601 mean) perform better than folds with more structurally divergent test proteins (fold~4: 0.525 mean).
Figure~\ref{fig:cv_violin} visualizes the per-fold and overall AUROC distributions.

\subsection{Feature Ablation}
\label{app:ablations}

Table~\ref{tab:feat_ablation} reports the effect of removing individual feature groups from the 28-dimensional node representation.
All experiments use the target-unseen split with consistent hyperparameters (LR=$10^{-3}$, 4 layers, hidden dim 128) for fair comparison.

\begin{table}[ht]
\centering
\caption{Feature ablation on target-unseen split. Delta ($\Delta$) relative to full model baseline. Positive $\Delta$ indicates improvement from removing the feature.}
\label{tab:feat_ablation}
\small
\begin{tabular}{@{}lcccc@{}}
\toprule
\textbf{Configuration} & \textbf{Val} & \textbf{Test} & \textbf{AUPRC} & \textbf{$\Delta$} \\
\midrule
Full model (baseline) & 0.584 & 0.477 & 0.406 & -- \\
$-$ pLDDT & 0.558 & 0.528 & 0.434 & +0.051 \\
$-$ SASA & 0.518 & 0.403 & 0.386 & $-$0.074 \\
$-$ Lysine indicator & 0.541 & \textbf{0.578} & \textbf{0.526} & +0.101 \\
$-$ Physicochemical & 0.580 & 0.551 & 0.447 & +0.074 \\
$-$ Disorder & 0.546 & 0.565 & 0.450 & +0.088 \\
Only AA one-hot & 0.529 & 0.532 & 0.437 & +0.055 \\
\bottomrule
\end{tabular}
\end{table}

Two findings stand out.
First, removing the lysine indicator produces the largest improvement (+0.101), confirming the paradox discussed in Section~\ref{sec:discussion}.
Second, SASA is the only feature whose removal consistently hurts performance ($-$0.074), confirming the importance of surface accessibility for degradability prediction---surface-exposed lysines are the functional targets for ubiquitin conjugation.
The ``only AA one-hot'' configuration (+0.055 vs.\ full) demonstrates that the amino acid identity alone provides substantial predictive signal, and additional features introduce noise in this small-data regime.

\begin{figure}[t]
\centering
\includegraphics[width=0.85\columnwidth]{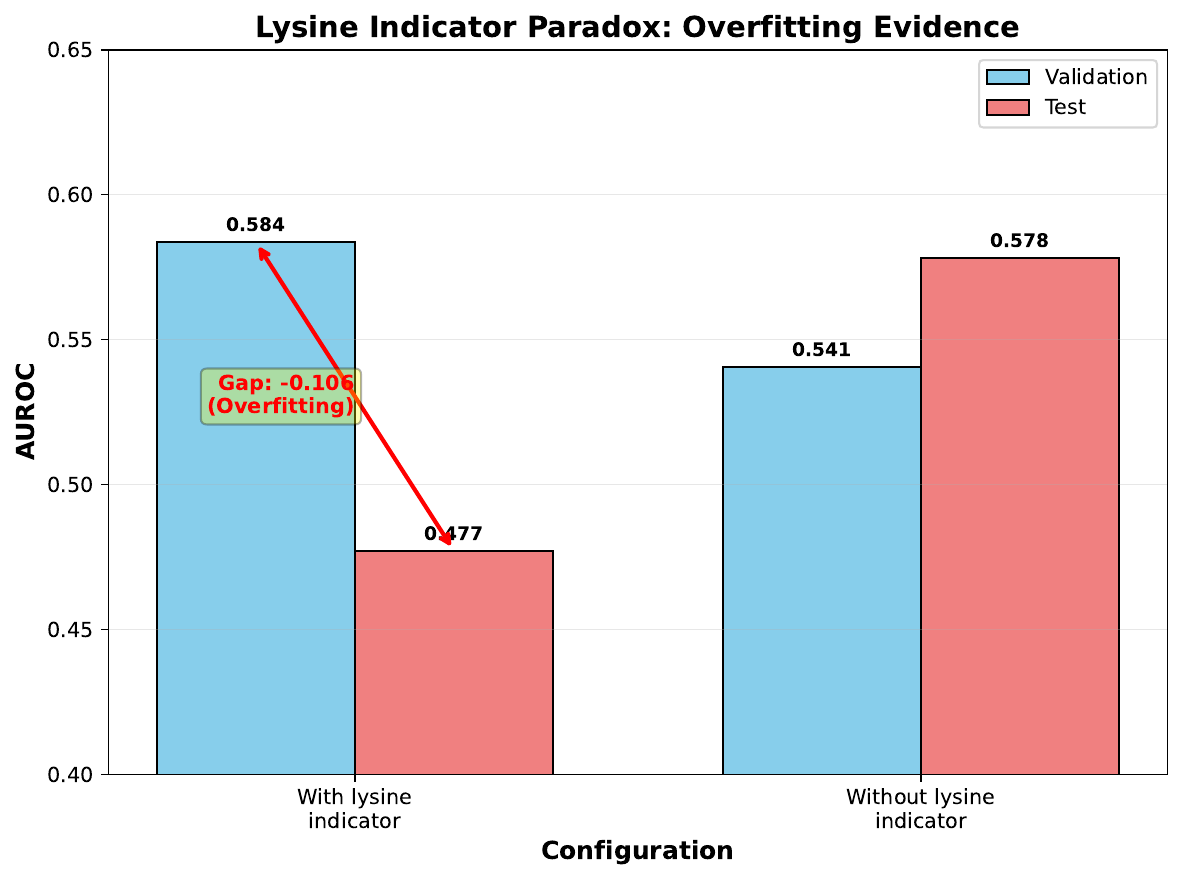}
\Description{Bar chart comparing validation and test AUROC with and without lysine indicator, showing overfitting pattern.}
\caption{Lysine indicator paradox: removing the lysine indicator improves test performance (+0.101) despite hurting validation ($-$0.043), revealing a classic overfitting pattern (val-test gap: $-$0.106 vs.\ +0.038).}
\label{fig:lysine_paradox}
\end{figure}

\subsection{Architecture Ablation}

Table~\ref{tab:arch_ablation} reports the effect of varying architectural hyperparameters on the target-unseen split.

\begin{table}[ht]
\centering
\caption{Architecture ablation (target-unseen AUROC). Default values marked with $\dagger$.}
\label{tab:arch_ablation}
\small
\begin{tabular}{@{}lcccr@{}}
\toprule
\textbf{Configuration} & \textbf{Val} & \textbf{Test} & \textbf{AUPRC} & \textbf{$\Delta$} \\
\midrule
\multicolumn{5}{@{}l}{\textit{GNN depth}} \\
\quad 2 layers & 0.527 & 0.466 & 0.434 & $-$0.011 \\
\quad 4 layers$^\dagger$ & 0.584 & 0.477 & 0.406 & -- \\
\quad 6 layers & 0.567 & \textbf{0.600} & \textbf{0.518} & +0.123 \\
\quad 8 layers & 0.577 & 0.565 & 0.465 & +0.088 \\
\midrule
\multicolumn{5}{@{}l}{\textit{Hidden dimension}} \\
\quad 64 & 0.557 & 0.446 & 0.406 & $-$0.031 \\
\quad 128$^\dagger$ & 0.584 & 0.477 & 0.406 & -- \\
\quad 256 & 0.568 & 0.582 & 0.502 & +0.105 \\
\midrule
\multicolumn{5}{@{}l}{\textit{Cross-attention heads}} \\
\quad 1 head & 0.574 & 0.524 & 0.433 & +0.047 \\
\quad 4 heads$^\dagger$ & 0.584 & 0.477 & 0.406 & -- \\
\quad 8 heads & 0.548 & 0.590 & 0.504 & +0.113 \\
\midrule
\multicolumn{5}{@{}l}{\textit{Radius cutoff}} \\
\quad 6\,\AA & 0.591 & 0.504 & 0.423 & +0.027 \\
\quad 8\,\AA$^\dagger$ & 0.584 & 0.477 & 0.406 & -- \\
\quad 12\,\AA & 0.540 & 0.414 & 0.384 & $-$0.063 \\
\bottomrule
\end{tabular}
\end{table}

Deeper networks (6 layers: +0.123), wider layers (256-dim: +0.105), and more attention heads (8: +0.113) all improve test AUROC, suggesting the default model is under-parameterized for this task.
However, these improvements are not reflected in validation AUROC (which sometimes decreases), underscoring the validation--test discordance discussed in Section~\ref{sec:discussion}.
The 12\,\AA{} radius cutoff hurts performance ($-$0.063), likely because the denser graph introduces noise from distant, structurally unrelated residues.

\subsection{Training Ablation}

\begin{table}[ht]
\centering
\caption{Training ablation (target-unseen AUROC). Default values marked with $\dagger$.}
\label{tab:train_ablation}
\small
\begin{tabular}{@{}lccr@{}}
\toprule
\textbf{Configuration} & \textbf{Val} & \textbf{Test} & \textbf{$\Delta$} \\
\midrule
\multicolumn{4}{@{}l}{\textit{Learning rate}} \\
\quad $5\times10^{-3}$ & 0.500 & 0.500 & +0.023 \\
\quad $10^{-3}$$^\dagger$ & 0.584 & 0.477 & -- \\
\quad $5\times10^{-4}$ & 0.554 & \textbf{0.666} & \textbf{+0.189} \\
\midrule
\multicolumn{4}{@{}l}{\textit{Dropout rate}} \\
\quad 0.0 & 0.578 & 0.616 & +0.139 \\
\quad 0.1$^\dagger$ & 0.584 & 0.477 & -- \\
\quad 0.2 & 0.558 & 0.523 & +0.046 \\
\bottomrule
\end{tabular}
\end{table}

The learning rate is the single most impactful hyperparameter: reducing from $10^{-3}$ to $5\times10^{-4}$ improves test AUROC by 18.9 percentage points (0.477$\to$0.666), the largest single improvement in all ablations.
This suggests the default LR causes the model to converge to a sharp minimum that generalizes poorly to unseen proteins, while the lower LR finds a flatter minimum with better generalization.
Removing dropout also helps substantially (+0.139), indicating that the model benefits from more capacity---regularization via dropout discards useful structural signals in this data-limited regime.

\subsection{Extended Classification Metrics}

Table~\ref{tab:detailed_class} consolidates confusion matrices and extended classification metrics across all three splits.

\begin{table}[ht]
\centering
\caption{Detailed classification metrics. Upper: confusion matrices at optimal F1 threshold. Lower: extended metrics. TP/FP/TN/FN = true/false positive/negative.}
\label{tab:detailed_class}
\scriptsize
\begin{tabular}{@{}l@{\,}r@{\,}r@{\,}r@{\,}r@{\;\,}c@{\;\,}c@{\;\,}c@{\;\,}c@{\;\,}c@{\;\,}c@{}}
\toprule
\textbf{Split} & \textbf{TP} & \textbf{FP} & \textbf{TN} & \textbf{FN} & \textbf{Thr.} & \textbf{AUC} & \textbf{F1} & \textbf{MCC} & \textbf{Prec.} & \textbf{Rec.} \\
\midrule
Tgt-uns. & 144 & 162 & 112 & 55 & 0.10 & .657 & .570 & .137 & .471 & .724 \\
E3-uns. & 248 & 110 & 585 & 163 & 0.45 & .811 & .645 & .460 & .693 & .603 \\
Random & 141 & 131 & 162 & 32 & 0.60 & .775 & .634 & .361 & .518 & .815 \\
\bottomrule
\end{tabular}
\end{table}

\begin{figure*}[t]
\centering
\includegraphics[width=0.9\textwidth]{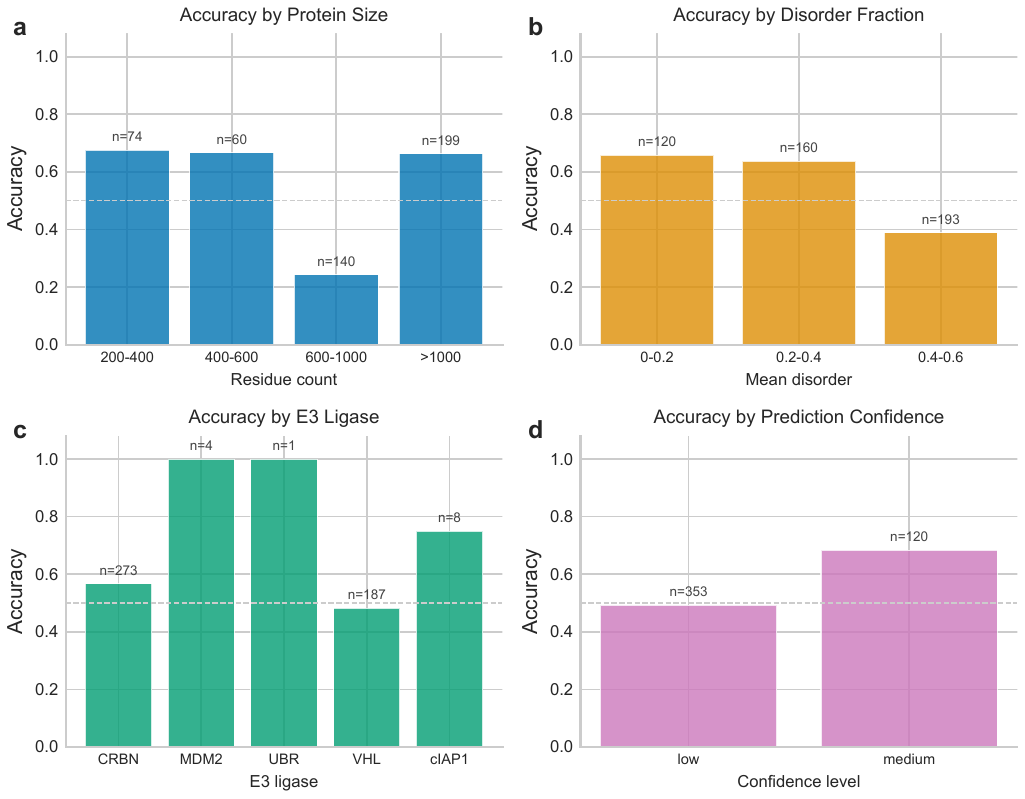}
\Description{Four-panel bar chart showing classification accuracy stratified by protein size, disorder fraction, E3 ligase, and prediction confidence on the target-unseen split, with sample counts above each bar.}
\caption{Stratified error analysis on the target-unseen split. (a)~Accuracy by protein size: large proteins (600--1000 residues) are hardest to classify. (b)~Accuracy by disorder fraction: highly disordered targets ($>$0.4) have lower accuracy. (c)~Accuracy by E3 ligase: CRBN outperforms VHL; rare E3s have too few samples for reliable estimates. (d)~Accuracy by prediction confidence: medium-confidence predictions are more accurate than low-confidence ones. Sample counts shown above each bar.}
\label{fig:error_analysis}
\end{figure*}

The confusion matrices reveal distinct operating characteristics across splits.
On the target-unseen split, the optimal threshold is very low (0.10), reflecting the model's tendency to assign moderate scores to unseen proteins.
The resulting high recall (0.724) at low precision (0.471) is appropriate for a screening application where missing a true degradable target is more costly than including a false positive.
The E3-unseen split operates at a balanced threshold (0.45) with the best precision (0.693), reflecting the model's confidence in cross-E3 predictions.
The low MCC (0.137) on target-unseen reflects the threshold-dependent nature of binary classification: AUROC and AUPRC, which evaluate the full ranking, paint a more favorable picture.
Figure~\ref{fig:error_analysis} provides a stratified error analysis across protein size, disorder fraction, E3 ligase, and prediction confidence.

\subsection{Calibration Metrics}
\label{app:calibration}

\begin{table}[ht]
\centering
\caption{Probability calibration metrics (extended). ECE = Expected Calibration Error, MCE = Maximum Calibration Error. ECE $<$ 0.05 = excellent; ECE $<$ 0.10 = good.}
\label{tab:calibration_app}
\small
\begin{tabular}{@{}lcccc@{}}
\toprule
\textbf{Split} & \textbf{Brier} & \textbf{Log Loss} & \textbf{ECE} & \textbf{MCE} \\
\midrule
Target-unseen & 0.240 & 0.674 & \textbf{0.029} & 0.135 \\
E3-unseen & \textbf{0.189} & \textbf{0.566} & 0.123 & 0.203 \\
Random & 0.219 & 0.628 & 0.165 & 0.379 \\
\bottomrule
\end{tabular}
\end{table}

\begin{figure}[t]
\centering
\includegraphics[width=0.75\columnwidth]{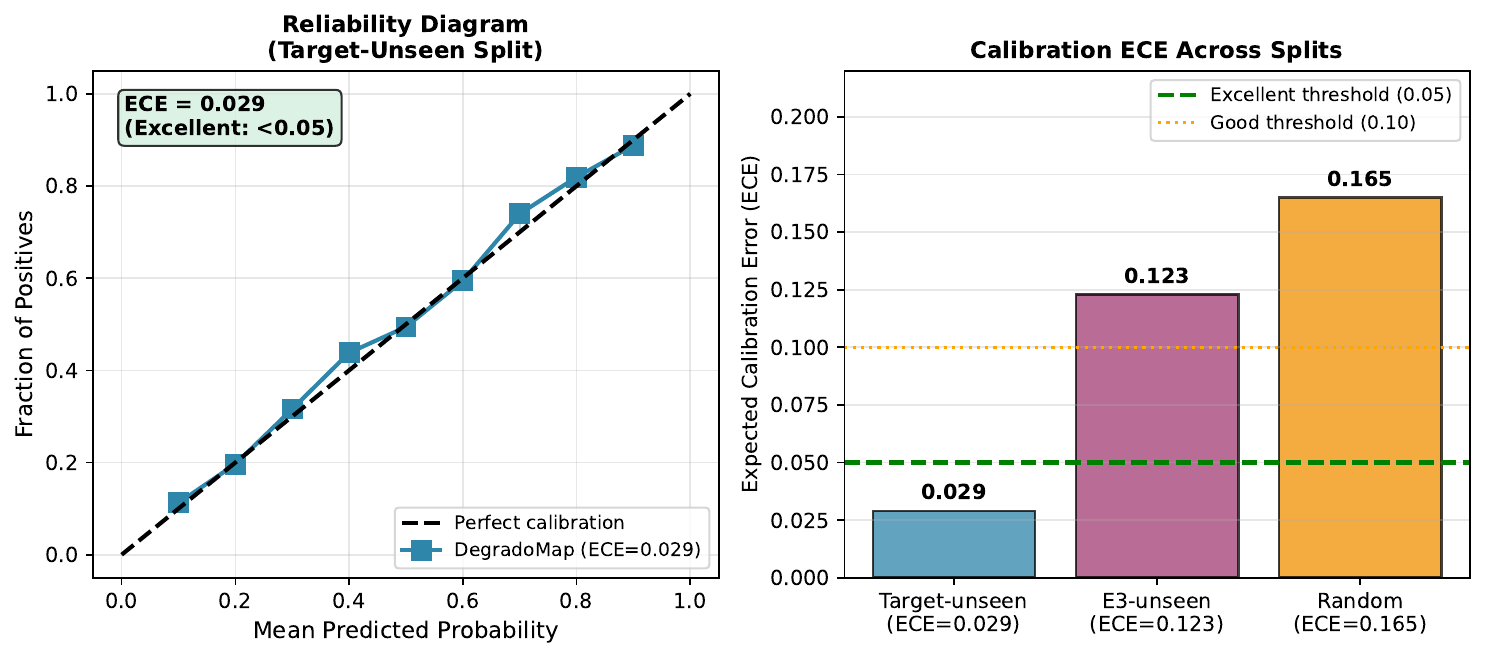}
\Description{Calibration curve plotting observed positive fraction versus mean predicted probability with a diagonal reference line representing perfect calibration, showing ECE and Brier score values.}
\caption{Calibration curve for the target-unseen split. Points show observed positive fraction vs.\ mean predicted probability in each bin; the diagonal represents perfect calibration. The model is well-calibrated (ECE = 0.029), with predicted probabilities closely tracking observed degradation rates.}
\label{fig:calibration}
\end{figure}

The target-unseen split shows excellent calibration (ECE = 0.029), indicating that predicted probabilities align well with observed degradation rates.
The E3-unseen split achieves the best Brier score (0.189) but higher ECE (0.123), suggesting some systematic overconfidence.
The random split shows the worst calibration (MCE = 0.379), likely due to distribution shift between validation-selected threshold and test data.
Figure~\ref{fig:calibration} shows the calibration curve for the target-unseen split.

\subsection{Precision@$k$ and NDCG Ranking Metrics}

\begin{table}[ht]
\centering
\caption{Precision@$k$, Recall@$k$, and NDCG@$k$ across evaluation splits.}
\label{tab:ranking}
\small
\begin{tabular}{@{}l cc cc cc@{}}
\toprule
& \multicolumn{2}{c}{\textbf{Target-uns.}} & \multicolumn{2}{c}{\textbf{E3-uns.}} & \multicolumn{2}{c}{\textbf{Random}} \\
\cmidrule(lr){2-3} \cmidrule(lr){4-5} \cmidrule(lr){6-7}
$k$ & P@$k$ & R@$k$ & P@$k$ & R@$k$ & P@$k$ & R@$k$ \\
\midrule
10  & 0.60 & 0.030 & 0.80 & 0.019 & 0.90 & 0.052 \\
20  & 0.70 & 0.070 & 0.90 & 0.044 & 0.95 & 0.110 \\
50  & 0.70 & 0.176 & 0.88 & 0.107 & 0.84 & 0.243 \\
100 & 0.69 & 0.347 & 0.86 & 0.209 & 0.75 & 0.434 \\
\midrule
& \multicolumn{2}{c}{\textbf{NDCG}} & \multicolumn{2}{c}{\textbf{NDCG}} & \multicolumn{2}{c}{\textbf{NDCG}} \\
\midrule
10  & \multicolumn{2}{c}{0.497} & \multicolumn{2}{c}{0.792} & \multicolumn{2}{c}{0.861} \\
20  & \multicolumn{2}{c}{0.602} & \multicolumn{2}{c}{0.866} & \multicolumn{2}{c}{0.910} \\
50  & \multicolumn{2}{c}{0.642} & \multicolumn{2}{c}{0.868} & \multicolumn{2}{c}{0.850} \\
\bottomrule
\end{tabular}
\end{table}

\begin{figure}[t]
\centering
\includegraphics[width=\columnwidth]{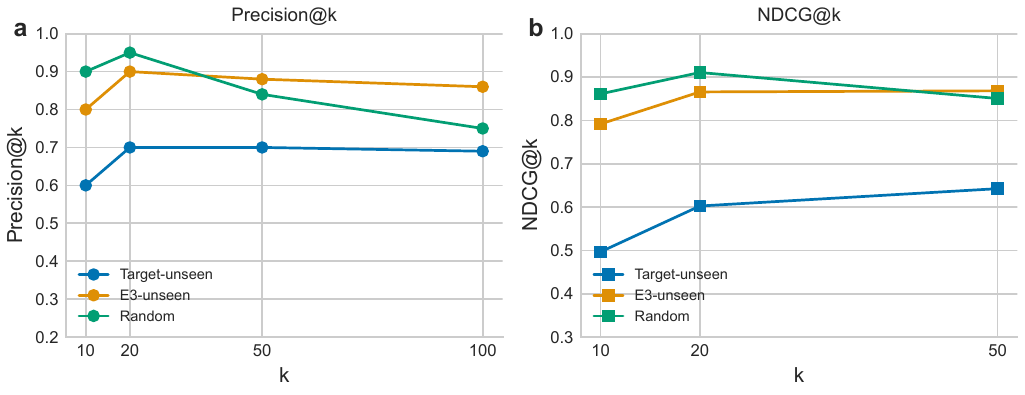}
\Description{Two-panel line plot showing Precision at k and NDCG at k trends across target-unseen, E3-unseen, and random evaluation splits for varying k values.}
\caption{Ranking metrics across evaluation splits. (a)~Precision@$k$ remains above 0.60 for target-unseen and above 0.80 for E3-unseen across all~$k$. (b)~NDCG@$k$ shows consistent ranking quality, with E3-unseen and random splits achieving $>$0.85 at $k = 20$.}
\label{fig:ranking}
\end{figure}

These ranking metrics are directly relevant to the screening use case, where a researcher would prioritize the top-$k$ predictions for experimental validation.
On the target-unseen split, P@20 = 0.70 means that 14 of the top 20 predicted degradable proteins are truly degradable---a practically useful hit rate for experimental follow-up.
E3-unseen ranking quality is particularly strong (P@20 = 0.90, NDCG@20 = 0.866).
Figure~\ref{fig:ranking} plots the Precision@$k$ and NDCG@$k$ trends.

\subsection{BRD4 Case Study}
\label{app:case_study}

BRD4 (Bromodomain-containing protein 4) is one of the most extensively studied PROTAC targets, with data available for multiple E3 ligases.
Table~\ref{tab:brd4} shows \ours{}'s predicted degradability scores alongside empirical degradation rates.

\begin{table}[ht]
\centering
\caption{BRD4 case study: predicted scores vs.\ empirical degradation rates across E3 ligases. Higher scores indicate predicted degradability.}
\label{tab:brd4}
\small
\begin{tabular}{@{}lcccc@{}}
\toprule
\textbf{E3 Ligase} & $n_{+}$ & $n_{-}$ & \textbf{Pred.\ Score} & \textbf{Emp.\ Rate} \\
\midrule
CRBN & 44 & 34 & 0.513 & 0.564 \\
VHL & 28 & 24 & 0.509 & 0.538 \\
FEM1B & 0 & 7 & 0.527 & 0.000 \\
MDM2 & 1 & 0 & 0.514 & 1.000 \\
KEAP1 & 1 & 0 & 0.527 & 1.000 \\
\bottomrule
\end{tabular}
\end{table}

The model assigns similar scores across E3 ligases, correctly reflecting BRD4's general degradability.
Predicted scores hover near 0.5, reflecting appropriate uncertainty when the PROTAC molecular structure is unknown.
The FEM1B case is instructive: despite 7 failed attempts (empirical rate 0.0), the model's score (0.527) is comparable to successful E3s, suggesting that FEM1B incompatibility reflects PROTAC design rather than intrinsic protein properties.

\subsection{Leave-One-E3-Out Cross-Validation}

\begin{table}[ht]
\centering
\caption{Leave-one-E3-out cross-validation results.}
\label{tab:leave_e3}
\small
\begin{tabular}{@{}lrrccc@{}}
\toprule
\textbf{E3 Held Out} & $n_{\text{test}}$ & $n_{+}$ & \textbf{AUROC} & \textbf{AUPRC} & \textbf{F1} \\
\midrule
VHL & 1,106 & 411 & 0.610 & 0.463 & 0.005 \\
CRBN & 1,871 & 737 & 0.606 & 0.496 & 0.000 \\
cIAP1 & 62 & 16 & 0.271 & 0.198 & 0.000 \\
\bottomrule
\end{tabular}
\end{table}

VHL and CRBN show consistent bidirectional transfer ($\sim$0.61 AUROC), validating the E3 compatibility module's ability to learn generalizable protein features.
The near-zero F1 scores indicate that the model's predicted probabilities cluster around 0.5 for held-out E3s (poor discrimination at any threshold), even though the ranking (AUROC) is meaningful.
cIAP1 performance is poor (0.271 AUROC) due to extreme sample scarcity---only 62 test samples with 16 positives.

\section{Dataset Details}
\label{app:dataset}

\subsection{E3 Ligase Distribution}

\begin{table}[ht]
\centering
\caption{E3 ligase distribution in PROTAC-8K dataset.}
\label{tab:e3_dist}
\small
\begin{tabular}{@{}lrr@{}}
\toprule
\textbf{E3 Ligase} & \textbf{Samples} & \textbf{\% Total} \\
\midrule
CRBN (Cereblon) & 2,011 & 62\% \\
VHL (von Hippel-Lindau) & 1,124 & 34\% \\
cIAP1 & $\sim$30 & 1\% \\
MDM2 & $\sim$20 & $<$1\% \\
XIAP & $\sim$15 & $<$1\% \\
DCAF1, KEAP1, FEM1B, others & $\sim$60 & 2\% \\
\bottomrule
\end{tabular}
\end{table}

The heavy skew toward CRBN and VHL reflects the current state of PROTAC research: these two E3 ligases have well-characterized ligands (thalidomide analogs for CRBN, VHL ligands based on hydroxyproline) and dominate clinical programs.
Novel E3 ligases (DCAF1, KEAP1, RNF114) are emerging in the literature but have minimal PROTAC data.

\subsection{Data Split Construction}

\begin{table}[ht]
\centering
\caption{Data split statistics.}
\label{tab:splits}
\small
\begin{tabular}{@{}lcccc@{}}
\toprule
\textbf{Split} & \textbf{Train} & \textbf{Val} & \textbf{Test} & \textbf{Description} \\
\midrule
Random & 2,170 & 465 & 466 & 70/15/15 random \\
Target-unseen & 2,218 & 410 & 473 & Protein-level grouping \\
E3-unseen & 1,643 & 352 & 1,106 & VHL held out entirely \\
\bottomrule
\end{tabular}
\end{table}

For the target-unseen split, we group samples by UniProt accession and assign entire protein groups to train, validation, or test.
This ensures strict protein-level separation: no data leakage from shared protein targets.
The E3-unseen split holds out all 1,106 VHL samples for testing, training only on CRBN-dominated data.

\subsection{Label Construction}

Binary degradation labels are constructed from quantitative experimental measurements:
\begin{itemize}
    \item \textbf{Positive (degrader)}: DC$_{50} < 1$\,$\mu$M AND D$_{\text{max}} > 50\%$
    \item \textbf{Negative (non-degrader)}: DC$_{50} \geq 1$\,$\mu$M OR D$_{\text{max}} \leq 50\%$, OR experimentally confirmed as inactive
\end{itemize}
where DC$_{50}$ is the concentration at which 50\% of target protein is degraded and D$_{\text{max}}$ is the maximum degradation achieved.
These thresholds follow conventions established in the DegradeMaster benchmark~\cite{liu2025degrademaster}.

\subsection{AlphaFold Structure Coverage}

AlphaFold-predicted structures (v6 API) were obtained for 152 of 155 unique protein targets.
Three proteins were unavailable in the human AlphaFold database: P03436 (Influenza A hemagglutinin), P0DTD1 (SARS-CoV-2 polyprotein Nsp3), and P36969 (GPX4, non-standard identifier).
Protein sizes range from 89 to 4,753 residues (median: 498).
Mean pLDDT across all structures was 79.3, with 89\% of residues having pLDDT $> 70$ (confident).

\subsection{External Validation Datasets}

We investigated three potential external validation datasets, none of which proved compatible:
the DeepPROTACs test set ($n = 16$) requires mol2 binding pocket files, the Bondeson kinase panel ($n \approx 52$) is behind a paywall, and PROTAC-PatentDB ($n = 63$K) lacks degradation labels.
This absence of compatible external validation reflects a broader challenge: different PROTAC prediction methods require fundamentally different input modalities, and label definitions vary across datasets.
Establishing a standardized benchmark with consistent evaluation protocols would significantly advance the field.

\subsection{Dataset Curation Considerations}

The PROTAC-8K dataset inherits several biases from the underlying PROTAC-DB 3.0:
\begin{itemize}
    \item \textbf{Publication bias}: Published PROTACs are enriched for successful degraders, as negative results are less frequently reported. The 39.4\% positive rate may underestimate the true failure rate of PROTAC attempts.
    \item \textbf{Target selection bias}: Extensively studied targets (BRD4, AR, ER) are overrepresented, while most human proteins have zero PROTAC data.
    \item \textbf{E3 ligase bias}: CRBN and VHL dominate (96\%) because they have well-characterized small-molecule ligands.
    \item \textbf{Assay heterogeneity}: DC$_{50}$ and D$_{\text{max}}$ values come from different labs using different assay conditions, cell lines, and incubation times, introducing measurement noise.
\end{itemize}
These biases should be considered when interpreting model performance and planning experimental validation.

\section{Model Architecture Details}
\label{app:architecture}

\subsection{Full Node Feature Specification}

\begin{table}[ht]
\centering
\caption{Complete 1,285-dimensional node feature breakdown for the improved model. Baseline uses AA one-hot (20-dim) instead of ESM-2, yielding 28 dimensions.}
\label{tab:nodefeats_full}
\scriptsize
\setlength{\tabcolsep}{2pt}
\begin{tabular}{@{}lr@{\;}l@{\;}l@{}}
\toprule
\textbf{Feature} & \textbf{Dim} & \textbf{Range} & \textbf{Description} \\
\midrule
ESM-2 embeddings & 1,280 & $\mathbb{R}$ & ESM-2-650M per-residue \\
Hydrophobicity & 1 & [$-$4.5, 4.5] & Kyte-Doolittle scale \\
Charge & 1 & \{$-$1, 0, 1\} & Formal charge at pH 7.0 \\
Size & 1 & [0, 1] & Normalized molecular weight \\
Polarity & 1 & \{0, 1\} & Binary polar/nonpolar \\
pLDDT & 1 & [0, 100] & AlphaFold confidence \\
SASA & 1 & [0, $\infty$) & Solvent-accessible surface area \\
Lysine flag & 1 & \{0, 1\} & Binary lysine indicator \\
Disorder & 1 & [0, 1] & Predicted disorder prob. \\
Known Ub site mask & 1 & \{0, 1\} & PhosphoSitePlus annotation \\
\midrule
\textbf{Total (improved)} & \textbf{1,285} & & \\
\textit{Total (baseline, AA one-hot)} & \textit{28} & & \\
\bottomrule
\end{tabular}
\end{table}

\subsection{Context Encoder Feature Groups}

The 59 cellular context features from DepMap are organized into 8 functional groups, each processed by a dedicated 2-layer MLP before concatenation:

\begin{table}[ht]
\centering
\caption{Cellular context feature groups (59 total dimensions).}
\label{tab:context_feats}
\small
\begin{tabular}{@{}lrl@{}}
\toprule
\textbf{Group} & \textbf{Dim} & \textbf{Source} \\
\midrule
Gene expression & 8 & RNA-seq (representative cell lines) \\
Gene effect & 8 & CRISPR dependency scores \\
Copy number & 8 & Genomic CNV \\
Protein expression & 8 & Mass spectrometry proteomics \\
Mutation frequency & 8 & WES/WGS \\
Metabolomics & 8 & Metabolic pathway activity \\
Drug sensitivity & 8 & PRISM compound response \\
Pathway membership & 3 & Curated pathway annotations \\
\midrule
\textbf{Total} & \textbf{59} & \\
\bottomrule
\end{tabular}
\end{table}

\subsection{Loss Function Details}

The multi-task loss combines three components with fixed weights:
\begin{align}
    \mathcal{L}_{\text{BCE}} &= -\frac{1}{N}\sum_{i=1}^{N}\left[y_i \log \hat{y}_i + (1-y_i)\log(1-\hat{y}_i)\right] \\
    \mathcal{L}_{\text{Huber}} &= \frac{1}{N}\sum_{i=1}^{N} \begin{cases} \frac{1}{2}(y_i^c - \hat{y}_i^c)^2 & |y_i^c - \hat{y}_i^c| \leq \delta \\ \delta(|y_i^c - \hat{y}_i^c| - \frac{1}{2}\delta) & \text{otherwise} \end{cases} \\
    \mathcal{L}_{\text{focal}} &= -\frac{1}{N}\sum_{i=1}^{N} \alpha_t (1-p_t)^\gamma \log(p_t)
\end{align}
with Huber threshold $\delta = 1.0$, focal $\gamma = 2.0$, and weights $\lambda_1 = 1.0$, $\lambda_2 = 0.5$, $\lambda_3 = 0.3$.
The focal loss component downweights well-classified examples, focusing training on hard cases near the decision boundary---important given the moderate class imbalance (39.4\% positive).

\section{Computational Efficiency}
\label{app:efficiency}

\subsection{Model Size and Speed Comparison}

\begin{table}[ht]
\centering
\caption{Computational efficiency comparison. Inference measured per protein (50 samples); training per epoch (100 samples). All on NVIDIA RTX 2080 Ti.}
\label{tab:efficiency}
\footnotesize
\begin{tabular}{@{}lr@{\;\;}c@{\;\;}c@{\;\;}c@{}}
\toprule
\textbf{Model} & \textbf{Params} & \textbf{Inf.\ (ms)} & \textbf{Train/ep.} & \textbf{Mem.} \\
\midrule
\ours{} (base) & 1.43M & 17.3\spm{0.6} & 5.19\,s & 75.7 \\
\ours{} (impr.) & 1.59M & -- & -- & -- \\
SchNet & 0.27M & 4.9\spm{0.0} & 1.40\,s & 57.7 \\
EGNN & 0.44M & 5.4\spm{0.3} & 1.70\,s & 60.9 \\
\bottomrule
\end{tabular}
\end{table}

\ours{} is $\sim$3.5$\times$ slower per inference than SchNet but achieves +15.2\% higher target-unseen AUROC.
At 17.3\,ms per protein, screening a library of 10,000 proteins requires $\sim$3 minutes, making \ours{} practical for high-throughput virtual screening applications.
Peak training memory (75.7\,MB) is well within the capacity of consumer GPUs, and the entire training pipeline completes in $\sim$26 hours on a single RTX 2080 Ti.

\subsection{Inference Scaling by Protein Size}

\begin{table}[ht]
\centering
\caption{Inference time by protein size for \ours{}.}
\label{tab:scaling}
\small
\begin{tabular}{@{}lccrc@{}}
\toprule
\textbf{Size} & \textbf{Residues} & \textbf{Mean res.} & $n$ & \textbf{Time (ms)} \\
\midrule
Small & $<$200 & 147 & 8 & 34.9\,\spm{11.6} \\
Medium & 200--500 & 381 & 62 & 33.5\,\spm{12.7} \\
Large & 500--1000 & 771 & 68 & 42.7\,\spm{19.0} \\
XLarge & $>$1000 & 1,430 & 33 & 39.9\,\spm{18.3} \\
\bottomrule
\end{tabular}
\end{table}

Inference time scales sub-linearly with protein size due to sparse radius graph construction: the number of edges grows proportionally to $N$ rather than $N^2$.
Even the largest proteins ($>$1000 residues, mean 1,430) require $<$60\,ms, with no special memory management needed.

\subsection{Training Resource Requirements}

\begin{table}[ht]
\centering
\caption{Full training pipeline resource requirements.}
\label{tab:resources}
\small
\begin{tabular}{@{}ll@{}}
\toprule
\textbf{Resource} & \textbf{Requirement} \\
\midrule
GPU & NVIDIA RTX 2080 Ti (11\,GB) \\
Total training time & $\sim$26 hours (20 epochs $\times$ 3 splits) \\
Per-epoch time & $\sim$5.2 seconds (full training set) \\
Peak GPU memory & 75.7\,MB \\
Data storage & $\sim$200\,MB (structures + features) \\
Model checkpoint & 5.5\,MB \\
\bottomrule
\end{tabular}
\end{table}

The modest computational requirements (single consumer GPU, $<$80\,MB memory) make \ours{} accessible to academic research groups without high-performance computing infrastructure.

\section{Reproducibility}
\label{app:reproducibility}

\subsection{Complete Hyperparameter Table}

\begin{table}[ht]
\centering
\caption{Complete hyperparameter specification for \ours{}. The improved model column shows tuned values; the baseline column shows original values where they differ.}
\label{tab:hyperparams}
\footnotesize
\begin{tabular}{@{}ll@{\;\;}c@{\;\;}c@{}}
\toprule
\textbf{Category} & \textbf{Hyperparameter} & \textbf{Impr.} & \textbf{Base.} \\
\midrule
\multirow{7}{*}{Arch.} & GNN layers & 4 & 4 \\
 & GNN hidden dim & 128 & 128 \\
 & GNN output dim & 64 & 64 \\
 & Node feat.\ dim & 1,285 & 28 \\
 & Cross-attn.\ layers & 2 & 2 \\
 & Attention heads & 4 & 4 \\
 & Radius cutoff & 8\,\AA & 8\,\AA \\
 & E3 embed.\ dim & 64 & 64 \\
 & Ctx MLP blocks & 3 & 3 \\
 & Ctx hidden dim & 128 & 128 \\
\midrule
\multirow{6}{*}{Training} & Optimizer & AdamW & AdamW \\
 & Learning rate & $5{\times}10^{-4}$ & $10^{-3}$ \\
 & $\beta_1, \beta_2$ & 0.9, 0.999 & 0.9, 0.999 \\
 & Weight decay & $10^{-2}$ & $10^{-2}$ \\
 & LR schedule & Cos.\ ann. & Cos.\ ann. \\
 & Batch size & 8 & 32 \\
 & Epochs (max) & 10 & 20 \\
 & Dropout & 0.05 & 0.1 \\
 & Early stopping & pat.=5 & -- \\
\midrule
\multirow{4}{*}{Loss} & $\lambda_{\text{BCE}}$ & 1.0 & 1.0 \\
 & $\lambda_{\text{Huber}}$ & 0.5 & 0.5 \\
 & $\lambda_{\text{focal}}$ & 0.3 & 0.3 \\
 & Focal $\gamma$ & 2.0 & 2.0 \\
\midrule
\multirow{2}{*}{ESM-2} & Model & 650M & -- \\
 & Embed.\ dim & 1,280 & -- \\
\bottomrule
\end{tabular}
\end{table}

\subsection{Random Seeds}

All experiments use seed 42 unless otherwise noted.
Cross-validation experiments (Section~\ref{app:cv}) use seeds \{42, 123, 456\} per fold to assess seed sensitivity.
Bootstrap confidence intervals use 1,000 iterations with sequential numpy random seeds.
Multi-seed GNN baseline experiments use seeds \{42, 123, 456\}.

\subsection{Software and Hardware}

\begin{table}[ht]
\centering
\caption{Software and hardware specifications.}
\label{tab:software}
\small
\begin{tabular}{@{}ll@{}}
\toprule
\textbf{Component} & \textbf{Version / Specification} \\
\midrule
\multicolumn{2}{@{}l}{\textit{Hardware}} \\
\quad GPU & NVIDIA GeForce RTX 2080 Ti (11\,GB) \\
\quad CPU & Intel Core i9 \\
\quad RAM & 32\,GB \\
\quad OS & Ubuntu Linux 5.15 \\
\midrule
\multicolumn{2}{@{}l}{\textit{Software}} \\
\quad Python & 3.13 \\
\quad PyTorch & 2.5.1+cu118 \\
\quad PyTorch Geometric & 2.4 \\
\quad scikit-learn & 1.3 \\
\quad NumPy & 1.26 \\
\quad Pandas & 2.0 \\
\bottomrule
\end{tabular}
\end{table}

\subsection{Data and Code Availability}

All data and code required for reproduction are publicly available:
\begin{itemize}
    \item \textbf{PROTAC-8K dataset}: \url{https://zenodo.org/records/14715718}
    \item \textbf{AlphaFold structures}: \url{https://alphafold.ebi.ac.uk/} (v6 API, human proteome)
    \item \textbf{DepMap features}: \url{https://depmap.org/portal/} (24Q2 release)
    \item \textbf{Code and pre-trained models}: \url{https://github.com/bryanc5864/DegradoMap}
\end{itemize}

The repository includes training scripts, evaluation pipelines, baseline implementations, and pre-trained model checkpoints for all three evaluation splits.
A Docker container is provided for exact environment reproduction.

\subsection{GNN Baseline Implementation Details}

All GNN baselines use identical graph construction (C$\alpha$ radius graph, 8\,\AA{} cutoff) and training procedures (AdamW, cosine annealing, 20 epochs) for fair comparison.
Table~\ref{tab:gnn_details} provides architecture specifications and complete multi-seed results.

\begin{table}[ht]
\centering
\caption{GNN baseline details. Upper: architecture specifications. Lower: multi-seed AUROC (3 seeds). High target-unseen variance indicates initialization sensitivity.}
\label{tab:gnn_details}
\scriptsize
\setlength{\tabcolsep}{2.5pt}
\begin{tabular}{@{}lrrll@{}}
\toprule
\textbf{Model} & \textbf{Params} & \textbf{Layers} & \textbf{Edge Feat.} & \textbf{Equivar.} \\
\midrule
\ours{} (base) & 1.43M & 4 & Distance & SE(3)-inv. \\
\ours{} (impr.) & 1.59M & 4 & Distance & SE(3)-inv. \\
E(3)-Equiv. & 1.04M & 4 & Vec.+dist. & E(3)-equiv. \\
SchNet & 0.27M & 4 & RBF exp. & SE(3)-inv. \\
EGNN & 0.44M & 4 & Distance & E($n$)-equiv. \\
\midrule
\textbf{Model} & \textbf{Split} & \textbf{Seed 42} & \textbf{Seed 123} & \textbf{Seed 456} \\
\midrule
\multirow{3}{*}{SchNet} & Random & 0.774 & 0.758 & 0.797 \\
& E3-unseen & 0.464 & 0.573 & 0.527 \\
& Target-uns. & -- & -- & 0.505 \\
\midrule
\multirow{3}{*}{EGNN} & Random & 0.662 & 0.706 & 0.769 \\
& E3-unseen & 0.592 & 0.628 & 0.474 \\
& Target-uns. & 0.641 & 0.453 & 0.460 \\
\bottomrule
\end{tabular}
\end{table}

SchNet uses 50 radial basis functions for continuous-filter convolutions.
EGNN updates both node features and coordinates at each layer, with the final coordinates discarded (only scalar features used for prediction).
The E(3)-equivariant variant of \ours{} replaces the invariant message passing with vector message passing and spherical harmonics edge encoding, while keeping the E3 compatibility and fusion modules identical.
EGNN shows particularly high variance on target-unseen (0.453--0.641), with seed 42 achieving performance close to \ours{} (0.641 vs.\ 0.657) while seeds 123 and 456 perform near random.
This instability suggests that EGNN's equivariant coordinate updates may be sensitive to initialization in the low-data regime.

\balance
\end{document}